\documentclass[preprint,12pt]{aastex}
\usepackage{natbib,epsfig}
\usepackage{graphicx}
\usepackage{multirow}
\usepackage{amsmath}
\tighten

\newcommand{\mt}{m_0}
\newcommand{\vt}{v_0}
\newcommand{\tdi}{t_0}
\newcommand{\bo}{breakout }
\newcommand{\lum}{luminosity }

\newcommand{\mh}{\widehat{m}}
\newcommand{\vh}{\widehat{v}}
\newcommand{\tauh}{\widehat{\tau}}
\newcommand{\dhat}{\widehat{d}}

\newcommand{\Eh}{\widehat{E}}
\newcommand{\Th}{\widehat{T}}
\newcommand{\Lh}{\widehat{L}}
\newcommand{\etah}{\widehat{\eta}}
\newcommand{\rh}{\widehat{r}}

\begin{document}

\title{Early supernovae light-curves following the shock-breakout}
\author{Ehud Nakar$^{1}$ and Re'em Sari$^{2}$ }
\affil{1. Raymond and Beverly Sackler School of Physics \&
Astronomy, Tel Aviv University, Tel Aviv 69978, Israel\\
2. Racah Institute for Physics, The Hebrew University, Jerusalem
91904, Israel\\}

\begin{abstract}
The first light from a supernova (SN) emerges once the SN shock
breaks out of the stellar surface. The first light, typically a UV
or X-ray flash, is followed by a broken power-law decay of the
luminosity generated by radiation that leaks out of the expanding
gas sphere. Motivated by recent detection of emission from very
early stages of several SNe, we revisit the theory of shock breakout
and the following emission, paying special attention to the
photon-gas coupling and deviations from thermal equilibrium. We
derive simple analytic light curves of SNe from various progenitors
at early times. We find that for more compact progenitors, white
dwarfs, Wolf-Rayet stars (WRs) and possibly more energetic
blue-supergiant explosions, the observed radiation is out of thermal
equilibrium at the breakout, during the planar phase (i.e., before
the expanding gas doubles its radius), and during the early
spherical phase. Therefore, during these phases we predict
significantly higher temperatures  than previous analysis that
assumed equilibrium. When thermal equilibrium prevails, we find the
location of the thermalization depth and its temporal evolution.
Our results are useful for interpretation of early SN light curves.
Some examples are: (i) Red supergiant SNe have an early bright peak
in optical and UV flux, less than an hour after breakout. It is
followed by a minimum at the end of the planar phase (about 10 hr),
before it peaks again once the temperature drops to the observed
frequency range. In contrast WRs show only the latter peak in
optical and UV. (ii) Bright X-ray flares are expected from all
core-collapse SNe types. (iii) The light curve and spectrum of the
initial breakout pulse holds information on the explosion geometry
and progenitor wind opacity. Its spectrum in more compact
progenitors shows a (non-thermal) power-law and its light curve may
reveal both the breakout diffusion time and the progenitor radius.
\end{abstract}


\section{Introduction}
A breakout of a shock through the stellar surface is predicted to be
the first electro-magnetic signal heralding the birth of a supernova
(SN)
\citep{Colgate74,Falk78,Klein78,Imshennik81,Ensman92,MatznerMcKee99}.
Before breakout the shock is propagating through the opaque stellar
envelope. The shock is radiation dominated (i.e., the energy density
behind the shock is dominated by radiation) and it accelerates while
propagating through the decreasing density profile of the envelope,
leaving behind the shock an expanding radiation dominated gas.
Following the shock breakout, photons continue to diffuse out of the
expanding stellar envelope producing a long lasting emission that
slowly decays with time
\citep[e.g.,][]{Grassberg71,Chevalier76,Chevalier92,Chevalier08,Piro09}.
The typical frequency of the breakout emission ranges from far
ultra-violet to soft $\gamma$-rays in core collapse SNe, and as we
show here, is in $\gamma$-rays in type Ia SNe. The typical frequency
of the following emission decreases to the visible-near UV bands
after a day. The energy released during the breakout increases with
the progenitor radius and can reach $\sim 0.1\%$ of the SN explosion
energy in a red supergiant. The luminosity of core collapse SNe
after a day is $\sim 10^{41}-10^{42} {\rm~ erg/s}$. Thus, the shock
breakout and the emission through the first day can be detected out
to the nearby Universe, but without any preceding knowledge of where
to look, their detection is challenging. Nevertheless, the search
worth the effort as this emission bears direct information on the
properties of the progenitor and the explosion, which are difficult
to obtain in any other way.

The development during the recent decade of sensitive UV, X-ray and
soft gamma-ray detectors, with relatively large fields of view, lead
to the discovery of several shock breakout candidates
\citep{Campana06,Soderberg08,Gezari08,Schawinski08,Modjaz09}.
Motivated by these, and by the  rising potential for future
detection of shock breakouts from various progenitors, we revisit
this topic. We develop an analytic model that provides light curves
(luminosity and temperature) starting from the breakout, through the
quasi-planar expansion phase to the spherical expansion phase, until
recombination and/or radioactive decay start playing a significant
role. These phases were explored in previous works, where the most
updated analytic study of the spherical phase was carried-out by
\cite{Chevalier92,Chevalier08}, and \cite{Waxman07,Rabinak10}. The
study of the planar phase was carried out only very recently by
\cite{Piro09} in the context of Type Ia shock breakout. The
advantage of our model is that we follow the photon-gas coupling
within the expanding gas. At each stage of the evolution we find the
location at which the observed temperature is determined. We find
whether the radiation at this place is in thermal equilibrium or not
and calculate the observed temperature. It turns out that the
temperature evolution during the planar phase and the early
spherical phase depends strongly on the thermal equilibrium of the
radiation just behind the shock in the breakout layer. In radiation
dominated shocks, radiation is out of thermal equilibrium if the
shock velocity is high enough \citep{Weaver76}, which is the case in
shock breakout from more compact progenitors and more energetic
explosions \citep{Katz09}. We provide the first calculation
(analytic or numerical) of the light curve in case that the observed
radiation is out of thermal equilibrium at the source. We also carry
out the first analytic calculation of the evolution of the location
of the thermalization depth, through the different phases, when the
radiation  is in thermal equilibrium. We use our model to examine
the light curve and spectrum of the initial pulse that is strongly
affected by light travel time effects and opacity of the progenitor
stellar wind. Finally, we use our model to explore the properties of
early SNe light curves resulting from various progenitor types
including red supergiants (RSG), blue supergiants (BSG), Wolf-Rayet
stars (WR) and white dwarfs (WD). We provide  simple formula of
early SN light curves for these different progenitors.

We present our model and its general results in section \ref{SEC
theory}. The bolometric luminosity and spectrum of the initial pulse
are discussed in section \ref{SEC initial pulse}. Early SNe light
curves resulting from various progenitors are presented in section
\ref{SEC lightcurve}. A reader that is interested only at the final
light curves should go directly to this section. In section \ref{SEC
comparison} we compare our calculations to previous analytic and
numerical studies. We summarize our main results in section \ref{SEC
Summary}. We ignore in this paper any cosmological redshift effects.

\section{Luminosity and observed spectrum following shock breakout}\label{SEC theory}
SN explosion drives a radiation dominated shock (i.e., the internal
energy in the shock downstream  is dominated by radiation) that
propagates and accelerates through the decreasing density profile of
the stellar envelope. The pre-explosion density profile near the
stelar radius, $R_*$, can be approximated by a power-law $\rho
\propto (R_*-r)^n$ where $r$ is the distance from the star center
and $n \approx 1.5-3$, depending on the stellar properties. In such
medium the blast-wave assumes a self similar profile in which the
shock velocity $v \propto \rho ^{-\mu}$, where $\mu$ is a very weak
function of $n$ and for the relevant range of $n$ can be well
approximated as a constant $\mu \approx 0.19$ \citep{Sakurai60}.
This purely hydrodynamic solution, which neglects heat conduction,
holds as long as the distance to the stellar edge, $R_*-r$, is
larger than the width of the radiation dominated shock, or
equivalently as long as the optical depth for photons to escape the
stellar envelope, $\tau(r)$, is larger than $c/v(r)$, where $c$ is
the light speed. Once $\tau \approx c/v$ the shock "breaks-out",
leaving behind an expanding hot gas-radiation sphere. We denote the
conditions at the front shell of gas at the breakout time (where
$\tau \approx c/v$ at the breakout) by subscript $_0$, and call this
shell {\it the \bo shell}. Given above assumptions our problem is
completely defined by the following parameters of the breakout shell
at the breakout time. The parameters of the inner gas layers are
directly related to it.
\begin{itemize}
\item $v_0$: the velocity of the shock when it breaks out.
\item $m_0$: the mass of the breakout shell.
\item $d_0$: the initial width of the breakout shell.
\item $R_*$: the radius of the star.
\item $n$: the power law index describing the pre-explosion density profile, typically $1.5 \le n \le 3$
\end{itemize}
Other characteristics of the breakout shell can be calculated from
the above. For example,
\begin{itemize}
\item $\tau_0=\kappa_T m_0/R_*^2$ is the initial optical depth for Thompson scattering, where $\kappa_T$ is the Thompson cross section per unit of mass.
\item $t_0=d_0/v_0$ is the initial dynamical time of the breakout shell. It is also its initial diffusion time.
\item $E_0=m_0v_0^2$ is the initial thermal energy of the breakout shell.
\end{itemize}

Conceptually, it is useful to treat the whole expanding envelope as
a series of successive shells. Any shell deeper than the breakout
shell, can be characterized by its mass $m>m_0$. \cite{Sakurai60}'s
hydrodynamic solution discussed above indicates that for these
deeper shells
\begin{itemize}
\item $v=v_0 (m/m_0)^{-0.19n \over n+1}$ is the velocity of the shell of mass $m$.
\item $d_i=d_0 (m/m_0)^{1 \over n+1}$ is the width of a shell of mass $m$ at shock breakout.
\item $E_i=mv^2=E_0 (m/m_0)^{1+0.62n \over n+1}$ is the internal energy of a shell of mass $m$ at shock breakout.
\end{itemize}
The subscript $_i$, for time dependent quantities, indicates initial
values.

Initially, after shock crossing but before significant expansion,
the thermal and kinetic energies in each shell are equal, hence the
expression for the initial internal energy given above. Following
shock breakout the stellar envelope expands and the width of each
shell increases as it accelerates on the expense of the gas thermal
energy. The acceleration ends once the shell width is multiplied
several times, approaching a coasting velocity which is about twice
its initial one \citep{MatznerMcKee99}. Hence, up to a factor of
order unity, the initial velocity right after the shock equals its
final coasting velocity, $v$. The internal energy on the other hand
falls as the volume of the shell increases.

Following shock breakout, photons continue to diffuse out of the
expanding stellar envelope. Below we derive the luminosity and
typical frequency of the diffusing photons. We focus on the early
phases, where the gas ionization fraction is high and the opacity is
dominated by Thomson scattering (e.g. over free-free absorption).
This assumption breaks as soon as the observed temperature falls
below about $1$ eV (the gas density at this point is
$10^{-10}-10^{-9} {\rm ~g/cm^3}$). We also neglect the drop in the
number of free electrons due to recombination of fully ionized He
(which takes place at a few eV), and assume that the energy
injection due to recombination and radioactive decay are negligible.
These conditions prevail from the shock breakout up until about a
day after the explosion.

First, we derive the observed bolometric luminosity as a function of
time. It is dictated by the hydrodynamics and Thompson scattering
and does not depend on the thermal coupling between the diffusing
radiation and the gas (at the stages of interest the shells are
coasting and the hydrodynamics is decoupled from the radiation).
Then, we discuss the radiation-gas coupling and derive the observed
spectrum, both when the diffusing radiation is in thermal
equilibrium and when it is not. We do not provide the radius of the
photosphere (i.e., $\tau = 1$) since it has no observable signature
in the specific system at hand. Instead we provide the radius and
mass of the location at which the luminosity is determined (see
definition below), $\rh$ and $\mh$ and the radius and mass of the
location at which the observed temperature (color) is determined,
$r_{cl}$ and $m_{cl}$.

\subsection{Luminosity}\label{SEC luminosity}
At any given time the observed luminosity is dominated by the
innermost shell out of which photons can effectively diffuse, i.e.
where the diffusion time is comparable to the time past since
breakout, which is also the dynamical time. The reason is that the
diffusion time from inner shells is longer and therefore their
photons are confined and cannot be seen,  while the radiation from
outer shells already escaped the expanding gas and were observed at
earlier times. The criterion that the diffusion time is comparable
to the dynamical time is $\tau \approx c/v$. Thus, the bolometric
luminosity at time $t$ is determined by the radiation energy in the
shell that satisfies $\tau(t) \approx c/v$. We call this shell {\it
the luminosity shell}, and denote $\widehat{x}$ as the parameter $x$
of this shell. The luminosity is given then by $\Eh/t$.

To follow the evolution, we measure the time $t$ from breakout. The
relevant parameters of a shell are the width $d$, velocity $v$,
internal energy $E$, optical depth $\tau$, and photon diffusion time
$t_d$, all of which, except for $v$, are time dependent ($v$ only
increases by a modest factor from its initial value, and is
approximated here as constant). The following relations holds for
each shell\footnote{The last approximation $d \approx vt$ holds at
$\tdi$ only for $\mt$. For other shells it holds at $t>d_i/v$.
However, these shells are irrelevant before that time so this
approximation is valid for our analysis} at all times $t>t_0$:
\begin{eqnarray}
  \tau & = & \kappa_T \frac{m}{r^2}\cr
    t_d & = & \frac{\tau d}{c}\cr
    E & =  & E_i \left(\frac{r^2 d}{R_*^2 d_i}\right)^{1/3}\cr
    d & = &  d_i+vt \approx vt \cr
    r & \approx & R_*+vt
\end{eqnarray}
Our equation for the internal energy above assumes adiabatic
expansion of a radiation dominated gas. It is therefore not valid
for outer shells where $t>t_d$ which cooled radiatively. It is
applicable from the \lum shell and inward.

The hydrodynamic evolution has two phases - a planar phase and a
spherical phase. The evolution of a shell is approximately planar as
long as its radius did not double and it is spherical at later
times. As we show below the \bo shell is also the \lum shell while
the evolution of the \bo shell is planar. Thus we separate the
temporal evolution into two asymptotic regimes, $(\tdi<) t \ll t_s$
and $t \gg t_s$, where $t_s=R_*/\vt$ is the time of transition
between the planar and spherical geometries of the breakout shell.

During the planar phase $r \approx R_*$, implying a constant $\tau$.
Since $d \propto t$ we obtain that $t_d/t$ is constant in time for
all the shells during their planar evolution. As we have seen above
at the beginning of the planar phase (i.e., $t=t_0$), the diffusion
time equals the dynamical time only for the \bo shell. Therefore,
the \bo shell is also the \lum shell (i.e., $\mh \approx \mt$)
throughout the planar phase. The implication is that throughout the
planar phase we continue to observe photons from roughly the \bo
shell. Adiabatic cooling then dictates
\begin{equation}\label{EQ Eplanar}
     \Eh(t<t_s) \approx  E_0
     (t/\tdi)^{-1/3}.
\end{equation}

During the spherical phase $r \approx vt$ and the opacity is,
 $ \tau \approx
 \tau_0 \frac{m}{\mt}\left(\frac{R_*}{vt}\right)^{2}=\tau_0 \left(\frac{m}{\mt}\right)^{\frac{1.38n+1}{n+1}}\left(\frac{t}{t_s}\right)^{-2}.$
The \lum shell satisfies, by definition, $\widehat{t_{d}} = t$,
which is equivalent to $\tau = c/v$. Combining these two equations
for $\tau$, we find that at $t>t_s$ the \lum shell evolves with $\mh
\propto t^{\frac{2(n+1)}{1.19n+1}}$, $\vh \propto
t^{-\frac{0.38n}{1.19n+1}}$, $\rh \propto
t^{\frac{0.81n+1}{1.19n+1}}$, $\dhat_i \propto
t^{\frac{2}{1.19n+1}}$ and $\tauh \propto
t^{\frac{0.38n}{1.19n+1}}$. Using these relations, and following the
adiabatic cooling of the shells we obtain
\begin{equation}\label{EQ Espherical}
     \Eh (t>t_s) \approx
 E_0\left(\frac{t_s}{\tdi}\right)^{-1/3}\left(\frac{t}{t_s}\right)^{\frac{1.29n+5}{3(1.19n+1)}}.
\end{equation}

The observed bolometric luminosity, given by $\Eh/t$, is therefore:
\begin{equation}\label{EQ Luminosity}
    L_{obs} \approx  \frac{E_0}{\tdi} \left\{\begin{array}{lr}
    \left(\frac{t}{\tdi}\right)^{-4/3} & ~~t<t_s \\
    \left(\frac{t_s}{\tdi}\right)^{-4/3}\left(\frac{t}{t_s}\right)^{-\frac{2.28n-2}{3(1.19n+1)}} & ~~t>t_s
    \end{array}\right .
\end{equation}
The ratio $E_0/t_0$ is simply the initial luminosity $L_0$. This
bolometric luminosity falls as $t^{-4/3}$ in the planar phase and
falls much more modestly as $t^{-0.17}$ to $t^{-0.35}$ (for $1.5\leq
n \leq 3$) during the spherical phase. The total energy released in
a logarithmic time unit, $Lt$, increases once the spherical phase
sets on. The reason is that most of the energy in the shock is left
behind it, in deeper more massive shells, as the shock accelerates
(indeed we have seen that $E_i$ is an increasing function of $m$).
Therefore inner shells, which were previously opaque, become
transparent during the spherical phase and those contain increasing
amount of energy.

The radius and mass of the \lum shell are:
\begin{equation}\label{EQ r_hat}
    \rh \approx  R_* \left\{\begin{array}{lr}
    1 & ~~t<t_s \\
    \left(\frac{t}{t_s}\right)^{\frac{0.81n+1}{1.19n+1}} & ~~t>t_s
    \end{array}\right. ,
\end{equation}

\begin{equation}\label{EQ m hat}
    \mh \approx  m_0 \left\{\begin{array}{lr}
    1 & ~~t<t_s \\
    \left(\frac{t}{t_s}\right)^{\frac{2(n+1)}{1.19n+1}} & ~~t>t_s
    \end{array}\right. .
\end{equation}

\subsection{Spectrum}\label{Sec: spectrum theory}
Throughout the early phases that we consider here, the radiation
dominates the heat capacity. Therefore, the luminosity derived above
is independent of the thermal coupling between the photons and the
gas (even a full coupling which equates the gas and radiation
temperatures everywhere has a negligible effect on the total
radiation energy). However, the optical depth of the \lum shell is
always much greater than unity ($\tauh_i=\tau_0 \gg 1$ and $\tauh$
only increases, or remains constant, with time). Thus, the escaping
photons would have many interactions with electrons as they cross
the envelope until finally escaping from the layer where $\tau=1$.
Therefore, while the total radiation energy is independent of the
photons-electrons interaction, the typical energy of each of the
escaping photons {\it is dictated} by their coupling to matter along
their way from the \lum shell outward.

At any given time the photon luminosity at $r>\rh$ is constant
(given by equation \ref{EQ Luminosity}). Therefore, there are two
extreme possibilities. (1) No radiation process can significantly
change the number of photons. Then, since photons dominate the heat
capacity, both luminosity and photon number flux are fixed
(independent of $r$ for $r>\rh$). Thus, the typical photon energy
must be fixed. The typical temperature is thus given by the photons
temperature $\widehat T$ of the  \lum shell. That temperature, in
turn, would be dictated by the limited ability of radiation
processes in the luminosity shell to create photons. (2) Some
radiation process can create new photons, which can share the energy
of the diffusing photons. Then the typical energy of the photons
decreases as they diffuse out. The typical temperature then falls
below the photons temperature of the \lum shell. If such processes
exist at some radius $r>\rh$, how low can the temperature drop?
Photons would only be created up until the radiation thermalizes.
Given that the luminosity is constant, and that the diffusion time
is $\tau d/c$, the photon energy density for all shells external to
the \lum shell up until optical depth of unity is
\begin{equation}\label{EQ epsilon}
    \epsilon~(1 \leq \tau \leq \tauh) \approx \frac{\Lh \tau}{c r^2}.
\end{equation}
The thermal equilibrium temperature given such energy density is
\begin{equation}\label{EQ TBB_def}
    T_{BB} \equiv (\epsilon/a)^{1/4},
\end{equation}
where $a$ is the radiation constant. The temperature will never drop
below $T_{BB}$ since then photons would be absorbed rather than
emitted. As we show below, shells that can generate enough photons
to affect the photons temperature on their diffusion out of the \lum
shell, will gain full thermal equilibrium, while shells that are out
of thermal equilibrium do not affect the radiation temperature at
all.

To clarify, we provide an alternative description of the two options
discussed above. As long as the radiation and the gas are in thermal
equilibrium (i.e, $\epsilon \propto T^4$), the photon number flux
increases with $r$. Thus, the gas at $r>\rh$ must generate photons
at a sufficient rate, and share the energy of the outgoing photons
with the generated photons, in order to keep the radiation in
thermal equilibrium temperature, $T_{BB}$. The observed temperature,
is determined then by the outermost shell (at $r \geq \rh$) that is
in thermal equilibrium. If, on the other hand, none of the shells at
$r \geq \rh$ is in thermal equilibrium then  $T_{obs}=\Th \gg
\Th_{BB}$. Note that we refer here to the typical photon energy as
the radiation temperature also when the radiation is out of thermal
equilibrium.

We note that the relation $L=4\pi \sigma r^2 T^4$, which is sometime
used in the context of shock breakout, is wrong at any radius if
$kT$ is the typical photon energy at radius $r$ ($\sigma$ is the
Stefan-Boltzmann constant). The reason is that this equation, which
relates luminosity, radius and temperature at the point where $\tau
\approx 1$, assumes that the radiation is in thermal equilibrium at
this point, while at the regime we discuss here it is never the
case. At early times the optical depth is dominated by Thomson
scattering, which conserves photon number. Therefore, as a typical
photon should be absorbed at least once in shells that can keep
thermal equilibrium, the shell at $\tau=1$ is always out of thermal
equilibrium and it plays no roll in determining neither the
temperature nor the luminosity.

Below we follow the photon-electron coupling in the different
regimes and determine the time dependent observed temperature as
function of initial conditions. We show that the observed
temperature can assume radically different evolution between cases
where the \bo shell is initially in equilibrium and cases where it
is not. We stress that we use the term observed temperature to
denote the typical energy of the observed photons also when the
radiation is out of thermal equilibrium and the spectrum is not a
blackbody. The observed temperature is often called color
temperature when radiation is in thermal equilibrium (to separate it
from the effective temperature, which we ignore throughout the paper
since it is not an observable). We denote the shell at which the
observed temperature is determined as the {\it color shell}. The
color shell is the thermalization depth when thermal equilibrium is
achieved, and as we show below the color shell is the \lum shell
when it is not.

\subsubsection{photon-electron coupling and thermal equilibrium}\label{SEC thermal coupling}

The main processes that typically couple the gas and the radiation,
at the physical conditions of interest, are free-free and bound-free
emission and absorbtion and Compton or inverse Compton scattering.
The importance of the free-free coupling, compared to bound-free
coupling, depends on the metalicity, where in low metalicity
environments the free-free process is more dominant while in high
metalicity it is the bound-free. We assume here that free-free
process dominate and discuss briefly the effects of bound-free
coupling later.

Consider an isolated shell with a radiation dominated energy density
$\epsilon$, which generates its own photons by free-free emission.
Now assume that initially, even though it is radiation dominated, it
does not have enough photons to be in thermal equilibrium. The
radiation temperature is $T>T_{BB}$ and the hot photons must
``cool"\footnote{We use here the term ``cooling" in the sense that
the hot photons give their energy to cooler photons in the process.
The total radiation energy is either constant, or reduced only by
adiabatic expansion.}, by sharing their energy with newly generated
photons, for the system to approach thermal equilibrium. This is
achieved by emission from electrons. The energy content of these
electrons is negligible, but we assume that they can constantly gain
energy from the photons an remain at the radiation temperature (this
assumption is justified in appendix B). In this way, the energy of
the existing photons is shared with that of the new photons that are
constantly generated by free-free emission. Then, given a
sufficiently long time, the number density of generated photons is
$n_{BB} \approx aT_{BB}^4/3kT_{BB}$ and thermal equilibrium is
achieved. Therefore, the time required to achieve thermal
equilibrium in the shell is roughly
$n_{BB}/\dot{n}_{ph,ff}(T_{BB})$, where $\dot{n}_{ph,ff}$ is the
rate, per unit volume, at which free-free emission generates photons
with energy $h\nu \approx 3 kT$.

In each shell that generates its own photons the time available to
obtain thermal equilibrium is the time photons are confined to the
shell, i.e., $\min\{t,t_d\}$. We, therefore, define a thermal
coupling coefficient in the expanding gas:
\begin{eqnarray}\label{EQ eta Def}
\nonumber
  \eta &\equiv&\frac{n_{BB}}{\min\{t,t_d\}~\dot{n}_{ph,ff}(T_{BB})} \\
   &\approx&  \frac{7 \cdot 10^{5} {\rm ~s~}}{\min\{t,t_d\}}
    \left(\frac{\rho}{10^{-10}{\rm g/cm^{3}}}\right)^{-2} \left(\frac{kT_{BB}}{100 eV}\right) ^{7/2}  .
\end{eqnarray}
where we approximate $\dot{n}_{ph,ff} \approx 3.5 \cdot 10^{36} {\rm
s^{-1}~cm^{-3}~} \rho^2 T^{-1/2}$. $\rho$ is mass density (in ${\rm
g~cm^{-3}}$) and $k$ is the Boltzmann constant. In the definition of
$\eta$ we do not include Comptonization of low energy photons, which
for the physical systems of interest becomes important only when the
radiation is out of thermal equilibrium (see below). $\eta$ of the
breakout shell at the initial time $t_0$ is then approximated by
\begin{equation}\label{EQ etat_i}
    \eta_0 \approx 0.2 \left(\frac{\vt}{10^4~{\rm km/s}} \right)^{15/4} \left(\frac{\rho_0}{10^{-9}~{\rm g/cm^{3}}}
    \right)^{-1/8}.
\end{equation}
We took here into account the fact that the shock compresses the gas
by a factor of $7$ and we used the relation $T_{BB,0} \approx
 (\rho_0 \vt^2/a)^{1/4}$ (see also \citealt{Katz09}). The criterion
for thermal equilibrium behind the shock at the breakout is $\eta_0
< 1$, which can be translated to a limit on the the breakout
velocity \citep{Weaver76,Katz09}: $v_0<15,000 {\rm ~km/s}$ with a
negligible dependence on $\rho_0$ (as $\rho_0^{1/30}$). In case that
bound-free emission dominates over free-free emission $\eta_0$ can
be smaller by up to an order of magnitude (see section \ref{SEC
BoundFree}).

For $\eta <1$, the radiation is in thermal equilibrium and the
temperature is $T_{BB}$. Due to the relation between absorbtion and
emission $\eta \approx 1$ is also the requirement that a photon with
$h\nu =3k T_{BB}$ is absorbed on average once by free-free process
during the available time. Thus, if $t_d \leq t$ then $\eta=1$ is
equivalent to $\tau_{abs}\tau \approx 1$ where
 $\tau_{abs}$ is the free-free absorption optical depth for photons
 with $h\nu=3kT_{BB}$.

If $\eta>1$ then free-free processes are not enough to couple the
electrons to the radiation in the shell. Moreover, not enough
photons are generated at $h\nu \sim 3kT$ to obtain thermal
equilibrium. However, if Compton scattering provides enough coupling
so the electrons follow the radiation temperature, then the
temperature $T$ is determined by the total number of photons that
are generated at $h\nu \sim 3kT$ or that are generated at lower
frequency, but can be Comptonized to an energy $3kT$. Then free-free
emission generates photons until the temperature
satisfies\footnote{Note that under the assumption that initially the
energy density is radiation dominated, the system will always be
driven towards this equilibrium since the ``cooling" time of the
system is shorter at higher temperatures ($\propto T^{-1/2}$).}
$\epsilon/3kT=\min\{t,t_{d}\}~\dot{n}_{ph,ff}(T)\xi(T)$, where
$\dot{n}_{ph,ff}$ accounts only for generation of photons at $h\nu
\approx 3kT$ while $\xi(T)$ account for photons that are generated
at lower frequencies and Comptonized to temperature $T$. Therefore,
$\dot{n}_{ph,ff}(T)\xi(T)$ is the total production rate of photons
that can equally share the energy in the shell. Since free-free
emission produces roughly a constant number of photons for every
decade of energy below $kT$, the Comptonization correction factor
$\xi$ is logarithmic. If no photons can be Comptonized then $\xi=1$,
while if Comptonization is important $\xi$ may be much larger than
unity. Writing this condition in terms of $\eta$ we find that if
$\eta>1$ then $T \xi(T)^2=T_{BB}\eta^2$.

The availability of a low energy photon for computerization is
discussed by \cite{Weaver76}. At the temperature and density range
of interest ($0.1-50$ keV and $\sim 10^{-8}-10^{-10} {\rm~ gr/
cm^3}$) it is determined by the requirement that the photon energy
can be doubled before it is re-absorbed by free-free process. The
ratio between $T$ and the lowest energy photon that can be
Comptonized  is then \citep{Weaver76}:
\begin{equation}\label{EQ ymax}
    y_{max} \equiv \frac{kT}{h\nu_{min}}=3  \left(\frac{\rho}{10^{-9}~{\rm g/cm^{-3}}}\right)^{-1/2} \left(\frac{T}{\rm 100
    eV}\right)^{9/4}
\end{equation}
Using the approximated rate, per unit volume per unit of $y \equiv
h\nu/kT$, at which free-free emission generates photons
\citep{Svensson84} we find that for $y_{max}>1$ the number of
generated photons that can be Comptonized to $kT$ is larger by a
factor $\int_{\frac{1}{y_{max}}}^1 (0.8-\ln[y])\frac{dy}{y}$ than
the number of photons that free-free emission generates at $\approx
kT$. Thus for $y_{max}>1$:
\begin{equation}\label{EQ xi}
  \xi(T) \approx \max\left\{1,\frac{1}{2}\ln[y_{max}]\left(1.6+\ln[y_{max}]\right)\right\}.
\end{equation}
If $y_{max}<1$ then inverse compton does not contribute to the
number of photons, implying $\xi=1$.  All together the logarithmic
correction to the number of photons, $\xi$, ranges between $\xi=1$
at $T \lesssim 100$ eV to $\xi \approx 100$ at $T = 50$ keV.
Therefore, since in the systems of interest $T_{BB} \lesssim 100$ eV
Comptonization becomes important, and cannot be neglected, only once
the radiation is out of thermal equilibrium.

In summary, an isolated shell would arrive to a temperature of:
\begin{equation}\label{EQ Teta1}
T=T_{BB}
\begin{cases}
\frac{\eta^2}{\xi(T)^2} & \eta>1  \\
1  & \eta<1
\end{cases}
\end{equation}
If $\eta<1$ the spectrum is a blackbody. When $\eta>1$ the spectrum
is an optically thin thermal free-free emission (with electrons at
temperature $T$), which is modified by Comptonization. It is
therefore a Wien spectrum (determined by the number of photons) at
high frequencies and thermal spectrum (with $T$) at low
frequencies\footnote{ The exact spectrum depends on the time it
takes cold photons to double their energy by Comptonizition, (not
calculated here).  If photons have no time to be comptonized then
the free-free emission spectrum is not modified. Otherwise a Wien
spectrum, $F_\nu \propto \nu^3 exp[-h\nu/kT]$, is observed in the
region where the spectrum is dominated by inverse compton. $F_\nu
\propto \nu^0$ at lower frequencies, where the spectrum is dominated
by free-free emission. At $\nu<\nu_{min}$ (defined in equation
\ref{EQ ymax}) the spectrum is dominated by self-absorbtion and
$F_\nu \propto \nu^2$.}.

Equation \ref{EQ Teta1} assumes that electrons and photons have the
same temperature through the whole evolution. This is clearly the
case after thermal equilibrium is obtained and while $\eta<1$, since
free-free emission and absorbtion keep the electrons and photons
tightly coupled. In case that photons are out of thermal equilibrium
they approach equilibrium by ``cooling" via Compton scattering. This
scattering is the heating source of the electrons, while the main
electron cooling source is free-free emission. Therefore equation
\ref{EQ Teta1} assumes that the electron cooling is the slower
process. In appendix B we show that this is generally true in the
cases of interest, and that it may be marginally violated only in a
small region of the relevant parameter phase space, in which case
the modifications are likely to be small. Therefore we assume that
electrons and photons have the same temperature in shells that
generate their own photons and that equation \ref{EQ Teta1} is
always valid in these shells.

Equation \ref{EQ Teta1} neglects relativistic effects (pair
production, relativistic bremsstrahlung, etc.) and is, therefore,
accurate only for $T \lesssim 50$ keV \citep{Weaver76,Svensson84}.
We do not calculate here the exact temperature when eq. \ref{EQ
Teta1} predicts higher temperature. In this case pair production
will result in a lower temperature than eq. \ref{EQ Teta1} predicts.
In SNe, where the velocities are at most mildly-relativistic, the
temperature should fall within the range of soft gamma-ray
detectors, $\sim 50-200$ keV, when pair production is important
\citep{Katz09}.

\subsubsection{Succession of Shells \& The observed photon energy}\label{SEC succession}
In \S\ref{SEC thermal coupling} we considered the evolution of the
photon temperature in an isolated shell whose energy density
$\epsilon$ is known. We have assumed that the initial number of
photons was small, so that the shell itself produced all its
photons. We now discuss the interaction between shells as the
photons diffuse out. A shell external to the \lum shell receives
photons from inner shells. It could, potentially, lower the typical
photon energy by sharing the energy of the incoming photons with
more photons that are produced in the shell.

The effect of a shell on the radiation temperature depends on its
thermal coupling coefficient:
\begin{equation}\label{EQ eta}
  \eta=\eta_0 \left(\frac{T_{BB}}{T_{BB,0}}\right)^{7/2} \left(\frac{\rho}{\rho_0}\right)^{-2} \left(\frac{\min\{t,t_d\}}{\tdi}\right)^{-1} .
\end{equation}
A shell cannot affect the observed temperature if it does not
generate more photons than those arriving from inner layers. Since
the energy flux, $L$, is independent of $r$ at $r>\rh$, the number
flux that each shell can generate is inversely proportional to the
typical photons energy it could generate, given by equation \ref{EQ
Teta1}. For $\eta<1$ the number flux that a shell can generate is
$L/(kT_{BB})$ while for $\eta>1$ it is
$L/(kT_{BB}\eta^{2}\xi^{-2})$. $T_{BB}$ is a decreasing function of
$r$, and therefore shells with $\eta<1$ increase the number flux
while bringing the photons temperature to $T_{BB}$. On the other
hand, as we show later, $T_{BB}\eta^2$ is an increasing function of
$r$ (the variation of the logarithmic factor $\xi$ between shells
can be neglected), and therefore, shells with $\eta>1$ cannot change
the number flux of photons arriving from inner shells, and cannot
modify the photon energies. Moreover, $\eta$ is an increasing
function of $r$. Therefore, if the \lum shell is out of thermal
equilibrium, i.e., $\etah>1$, then all external shells have
$\eta(m<\mh)>1$. It follows, that the only shell which affects the
temperature when $\etah >1$ is the \lum shell, which is generating
its own photons.

The observed temperature is therefore:
\begin{equation}\label{EQ T_eta}
    T_{obs}= \left\{\begin{array}{lr}
                 \frac{\etah_c^2}{\widehat{\xi}_c^2}\; \Th_{BB} & \etah>1 \\
                 T_{BB}(\eta=1) & \etah<1
              \end{array}\right.
\end{equation}
Where $\etah_c$ and $\widehat{\xi}_c$ are calculated for the shell
that is currently the \lum shell, at the point where the photon
number in the shell was determined. Note that in case that $\etah<1$
the color shell is exterior to the \lum shell, where $t_d \leq t$.
Therefore the criterion $\eta=1$ is equivalent to $\tau_{abs}\tau
\approx 1$, which is a criterion for departure from thermal
equilibrium used in stellar envelope calculations, and in some of
the works on shock breakouts \citep[e.g.,][]{Ensman92}.

$T_{obs}$ is the typical photon energy of the observed radiation,
but the observed spectrum is not necessarily a blackbody. The
spectrum of the radiation at the color shell is determined only by
itself and is therefore that of an isolated shell with temperature
$T_{obs}$ (see the description below equation \ref{EQ Teta1}). As
discussed above, outer shells do not affect the number of photons
with energy  $h\nu \gtrsim kT_{obs}$, however they may contribute to
the number of photons with energy $h\nu < kT_{obs}$, thereby
modifying the low energy spectrum. We do not calculate here this
modification.

In the following sections we calculate the observed temperature by
first finding $T_{BB}$ in each shell. For shells external to the
\lum shell it is found using equation \ref{EQ Luminosity} and the
relation $\epsilon \approx L \tau/c r^2 $. For inner shells it is
determined by the adiabatic cooling during the expansion following
the shock passage. Then we use equation \ref{EQ eta Def} to
calculate $\eta$ and equation \ref{EQ T_eta} to find $T_{obs}$. We
define the color radius to be the radius at which the observed
temperature is determined (i.e., the radius of teh color shell),
namely $r_{cl}=\max\{\rh,r(\eta=1)\}$. We define $m_{cl}$ as the
color shell mass. i.e., $m(r>r_{cl})$. Note that we do not discuss
the radius of the photosphere where $\tau=1$, since it has no
observable consequence or physical importance in the context of
early SN light curves.

\subsubsection{The temperature during the planar phase ($\tdi < t  < t_s$)}

As discussed above (section \ref{SEC luminosity}), the optical depth
of each shell is fixed during the planar evolution and the breakout
shell is also the luminosity shell. Therefore, the observed
temperature can only be determined by the \bo shell or by shells
that are further out (i.e., $m\leq m_0$). Unfortunately, the
hydrodynamic evolution of these shells is uncertain, as it depends
on the poorly understood details of the radiation dominated shock
front as it breaks out of the envelope. As a working assumption, we
take the density profile at $m<\mt$ during the whole planar phase to
be similar to the pre-shock profile, i.e., $\rho =\rho_i d_0/\vt t$.
We verify later that other reasonable profiles do not strongly
affect the conclusions. Having $\rho(m,t)$ we use equation \ref{EQ
epsilon} to find $T_{BB}(m,t)$ and then equation \ref{EQ eta}
(noting that here $t_d \leq t$) to obtain:
\begin{equation}\label{EQ eta planar}
    \eta(m \leq \mt) = \eta_0
    \left(\frac{m}{\mt}\right)^{-\frac{17n+9}{8(n+1)}}
    \left(\frac{t}{\tdi}\right)^{-1/6}
\end{equation}
Thus, the thermal coupling increases slowly, but monotonically, with
mass and time, everywhere. This is true for any profile where the
density drops with radius. The number of photons in the \lum shell
increases continuously if it is out of thermal equilibrium (i.e.,
$\etah$ decreases with time so $\etah_c=\etah(t)$), and following
equations \ref{EQ T_eta} and \ref{EQ eta planar} the observed
temperature at $\tdi<t<t_s$ is:
\begin{equation}\label{EQ Tobs planar}
    T_{obs}= \left\{\begin{array}{lc}
                     T_{obs,0} \left(\frac{\widehat{\xi}(T_{obs})}{\xi_0(T_{obs,0})}\right)^{-2}\left(\frac{t}{\tdi}\right)^{-2/3} & \etah>1 ~(t < t_0 \eta_0^6)  \\
                     T_{BB,0} \eta_0^{~\frac{2(n+1)}{17n+9}} \left(\frac{t}{\tdi}\right)^{-\frac{2}{3}\frac{9n+5}{17n+9}}& \etah<1 ~(t > t_0
                     \eta_0^6)
                   \end{array}\right. .
\end{equation}
The value of $T_{obs,0}$ when $\eta_0>1$ is calculated by solving
the implicit equation \ref{EQ Teta1} for $T$ using $\eta_0$ and
$\xi_0(T_{obs,0})$. Note that for $\etah>1$ equation \ref{EQ Tobs
planar} is implicit due to the dependence of $\widehat{\xi}$ on
$T_{obs}$. We solve for $\widehat{\xi}$ by plugging $\rho=\rho_0
(\tdi/t)$ in equation \ref{EQ ymax}. $\widehat{\xi}$ is continuously
decreasing during the planar phase, but its decay is not described
by a power-law and so is the decay of $T_{obs}$. Nevertheless the
temperature instantaneous power-law decay index $\alpha=d
\ln(T_{obs})/d\ln(t)$ is bounded between that of an adiabatic
cooling and that of a constant $\xi$, i.e., $1/3< \alpha < 2/3$. The
drop in $\xi$ is smaller when $\eta$ is larger, implying that
$\alpha$ decreases with $\eta$.

The radius of the color shell is roughly constant during the planar
phase, $r_{cl} \approx R_*$. If $\eta_0<1$ then $m_{cl}$ depend
strongly on the unknown density profile at $m<\mt$. If $\eta_0>1$
the photospheric mass is constant $m_{cl}=\mt$. We find that $T_{BB}
\eta^{\;2}$ increases with the radius, in agreement with our
assertion that shells external to the \bo shell that are out of
thermal equilibrium, do not affect $T_{obs}$ since they generate
less photons than those arriving from inner shells. Finally, we
verify that taking other density profiles external to the breakout
shell such as a steeper decreasing density ($\rho=\rho_i d_i/\vt t$)
and a constant (in radius) density ($\rho=\rho_0 d_0/\vt t$) do not
significantly change equation \ref{EQ Tobs planar} (the power-law
indices when $\etah<1$ are changed by less than 0.1).

\subsubsection{The temperature during the spherical phase $t > t_s$}

The evolution of the spectrum during the early spherical phase
depends strongly on whether the \bo shell is in thermal equilibrium
at $t_s$. When it is, the color shell at $t_s$ is at $m<\mt$ and it
quickly moves inward (in Lagrangian sense) to $m>\mt$. Following
similar steps to those we used to derive equation \ref{EQ eta
planar} taking $\rho(\mt \leq m,t)$ we find
\begin{equation}
    \eta(\mt \leq m) =\eta_0 \left(\frac{t_s}{\tdi}\right)^{-1/6}
    \left(\frac{t}{t_s}\right)^\frac{42n+49}{12(1.19n+1)}
    \left(\frac{m}{\mt}\right)^{-\frac{22.32n+17}{8(n+1)}}.
\end{equation}
We consider here only $m<\mh$ where the time available for
equilibrium is $t_d$. The color shell is at $m(\eta=1)$ and the
observed temperature is\footnote{ we neglect here the slightly
different time evolution between $t_s$ and $t(\eta_0=1)$, which
introduces a negligible correction during this time.}:
\begin{eqnarray}\label{EQ Tobs spherical small etat}
\nonumber
  T_{obs}(\etah<1) &=& T_{BB,0} \eta_0^{~\frac{2(1.76n+1)}{22.32n+17}}
    \left(\frac{t_s}{\tdi}\right)^{-\frac{8.03n+6}{22.32n+17}} \\
   && \left(\frac{t}{t_s}\right)^{-\frac{18.48n^2+20.69n+6}{(1.19n+1)(22.32n+17)}}.
\end{eqnarray}
The color radius and the mass above it are
\begin{equation}\label{EQ rph}
r_{cl}(\etah<1)\approx R_*
\left(\frac{t}{t_s}\right)^{\frac{5.31n^2+9.09n+4.25}{(1.19n+1)(5.58n+4.25)}}
\end{equation}
and
\begin{eqnarray}\label{EQ mph}
\nonumber
  m_{cl}(\etah<1) & = & \mt~ \eta_0^\frac{2(n+1)}{5.58n+4.25}
  \left(\frac{t_s}{\tdi}\right)^{-\frac{n+1}{3(5.58n+4.25)}} \\
   &&\left(\frac{t}{t_s}\right)^{\frac{7(6n+7)(n+1)}{6(1.19n+1)(5.58n+4.25)}}.
\end{eqnarray}

\begin{figure}[h!]
\centerline{\includegraphics[width=8cm]{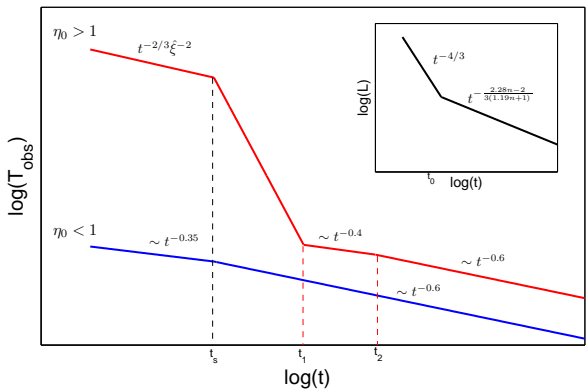}} \caption{{\it
Main:} A Schematic plot of the temperature evolution in the cases of
$\eta_0>1$ {\it (red line)} and  $\eta_0>1$ {\it (blue line)}. The
power-law indices indicated in the plot depend weakly on $n$ (see
the text) and the values given here are very good approximations
when $n$ ranges between $1.5$ and $3$. {\it Inset:} A Schematic plot
of the bolometric luminosity evolution. The luminosity, unlike the
observed temperature, is independent of the thermal coupling and the
value of $\eta$. } \label{fig1}
\end{figure}

The optical depth at the color shell decreases with time roughly as
$\tau_{cl} \propto m_{cl}^2/r_{cl} \propto t^{-0.4}$. Once
$\tau_{cl} = 1$ the opacity is not dominated anymore by Thomson
scattering and our model assumptions break. Typically, however, the
model assumptions break first by the temperature droping to $1$ eV
(and recombination becoming important) while $\tau_{cl}>1$. Note
that $r_{cl}(\etah<1)$ increases faster than $\rh$ with time. Thus,
if the \lum shell is at thermal equilibrium at a given time it will
remain in equilibrium at any later time. Note also that equations
\ref{EQ Tobs spherical small etat}-\ref{EQ mph} are valid whenever
$\etah<1$, regardless of the value of $\etah$ at earlier times. A
schematic light curve of this case is depicted in figure \ref{fig1}.

A very different evolution of $T_{obs}$ takes place if the \lum
shell is out of thermal equilibrium at $t_s$. Here evolution at the
beginning of the spherical phase is determined by the fact that the
thermal coupling of each shell increases (i.e., $\eta$ decreases)
while its evolution is quasi-planar (i.e. $t<R_*/v$). On the other
hand the thermal coupling becomes weaker during the spherical phase.
Therefore, the photon production of shells that are out of
equilibrium (i.e., $\eta
> 1$) is negligible during the spherical phase and the number of photons
in such shells  is determined by $\eta_c=\eta_{min}$, i.e. the
minimal value of $\eta$ obtained when the shell just doubled its
radius, and its thickness increased from the initial width $d_i$ to
$R_*$ . There, using equation \ref{EQ eta}, with
$(T_{BB}/T_{obs,0})^4 \approx (d_i/R_*)^{4/3}mv^2/m_0v_0^2$ and
$\rho/\rho_0 \approx mt_0/m_0t_s$, we obtain (noting that during the
planar evolution for the relevant shells $t \leq t_d$),
\begin{equation}\label{EQ eta_min}
\eta_{min}=\eta(t=R_*/v)=\eta_0\left(\frac{\tdi}{t_s}\right)^{1/6}
\left(\frac{m}{\mt}\right)^{-\frac{9.88n+5}{6(n+1)}}.
\end{equation}
Thus, the observed temperature drops very quickly at the beginning
of the spherical phase until 
$t_1 \equiv t(\etah_{min}=1)$:
\begin{equation}\label{EQ t1}
t_1=t_s
\left[\eta_0\left(\frac{\tdi}{t_s}\right)^{1/6}\right]^\frac{3(1.19n+1)}{9.88n+5}
\end{equation}
Between $t_s$ and $t_1$ the value of $\widehat{\xi}_c$ is also
dropping significantly until it becomes of order unity at $t_1$.
Therefore $T_{obs}$  does not evolve as a power-law during this
time, and we do not provide here a formula for its evolution.
Nevertheless the drop in $T$ is very quick ($\propto t^{-5.3}$ when
the evolution of $\widehat{\xi}_c$ is neglected) providing a strong
observational signature, regardless of the exact decay rate.

At $t_1$ the observed radiation is in thermal equilibrium (i.e.,
$T_{obs}=\Th_{BB}$), although $\etah
> 1$, since the radiation was in thermal equilibrium at $t=R_*/\vh$
and equilibrium is kept through adiabatic cooling. The observed
temperature during this phase, $\etah_{min}<1<\etah(t)$, is (note
that $\widehat{\xi}_c=1$ at this time):
\begin{equation}\label{EQ Tobs spherical t2}
    T_{obs}(t_1<t<t_2) = T_{BB,0}~ \eta_0^2 \left(\frac{t_s}{\tdi}\right)^{-2/3}
    \left(\frac{t_1}{t_s}\right)^{-\frac{21.27n+11}{3(1.19n+1)}}
    \left(\frac{t}{t_1}\right)^{-\frac{3n+2}{6(1.19n+1)}}
\end{equation}
The relation $T_{obs}=\Th_{BB}$ holds until $t_2 \equiv t(\etah=1)$.
Since $\etah \propto t^{\frac{12.48n+1}{6(1.19n+1})}$ we find:
\begin{equation}\label{EQ t2}
t_2=t_s
\left[\eta_0\left(\frac{\tdi}{t_s}\right)^{1/6}\right]^\frac{6(1.19n+1)}{12.48n+1}
.
\end{equation}
During the whole time that $\etah>1$ the color shell is also the
\lum shell ($r_{cl}=\rh$) and therefore:
\begin{equation}\label{EQ rpht1}
r_{cl}(t_s<t<t_2)=R_*
\left(\frac{t}{t_s}\right)^{\frac{0.81n+1}{1.19n+1}}
\end{equation}
and
\begin{equation}\label{EQ rpht1}
m_{cl}(t_s<t<t_2)=\mt~
\left(\frac{t}{t_s}\right)^{\frac{2(n+1)}{1.19n+1}}.
\end{equation}

At $t>t_2$ the color shell is exterior to the \lum shell
($\rh<r_{cl}$) and is in thermal equilibrium. Thus $T_{obs}$,
$r_{cl}$ and $m_{cl}$ at $t>t_2$ are given by equations \ref{EQ Tobs
spherical small etat}-\ref{EQ mph}. Finally, In all the different
regimes of the spherical phase $T_{BB} \eta^{\;2}$ increases with
the radius at $r>\rh$, implying that shells which are out of thermal
equilibrium in this radii range do not affect $T_{obs}$.  A
schematic light curve in case that the \lum shell is out of thermal
equilibrium at $t_s$ is depicted in figure \ref{fig1}.

\begin{figure}[h!]
\centerline{\includegraphics[width=8cm]{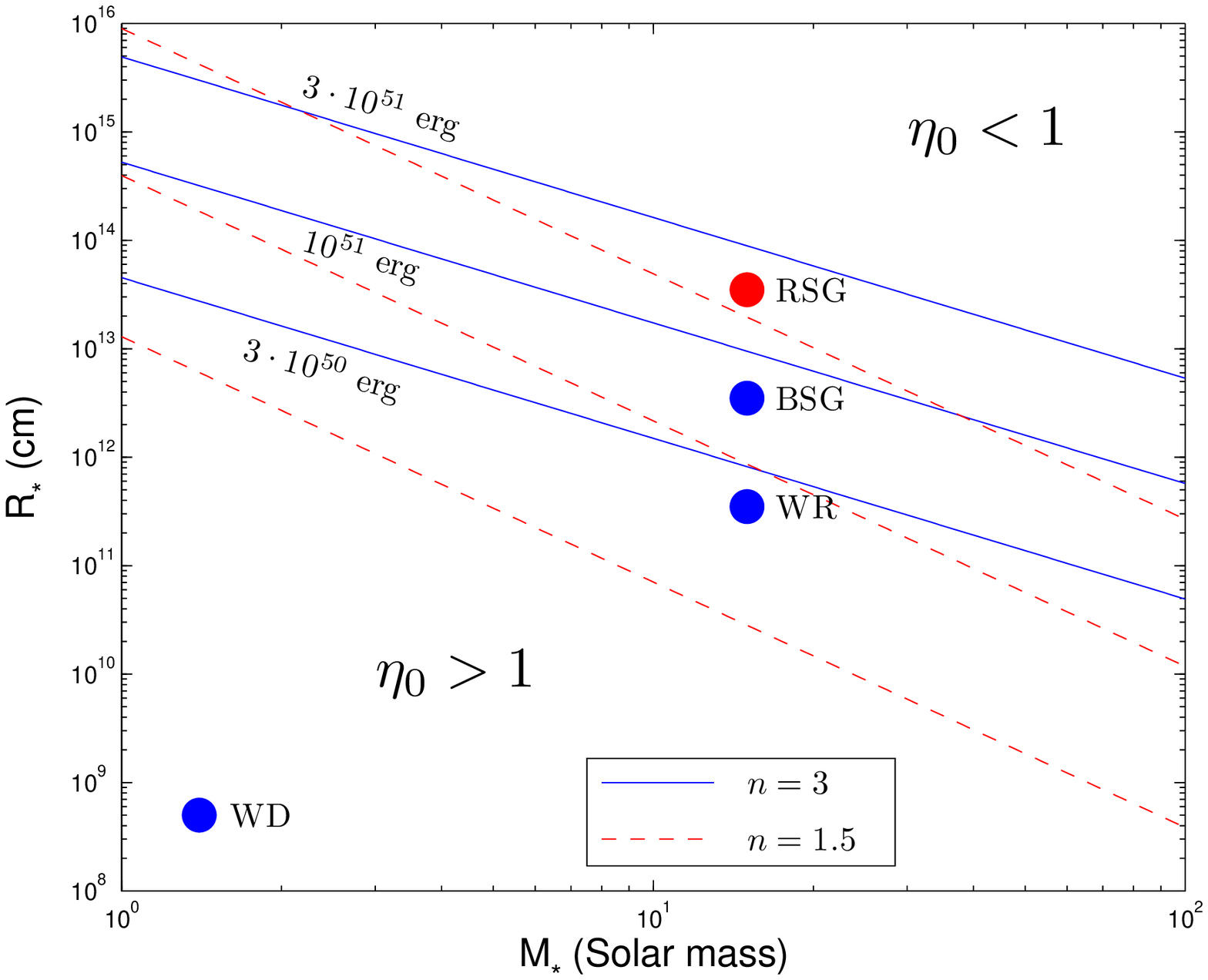}} \caption{{\it
Main:} Contours of $\eta_0=1$ in the $M_*$-$R_*$ plane for three
different values of the explosion energy ($3 \cdot 10^{50}$ erg,
$10^{51}$ erg and $3 \cdot 10^{51}$ erg), and two values of $n$
($n=3$ [{\it solid line}] and $n=1.5$ [{\it dashed line}]). We used
$\kappa = 0.34 {\rm ~cm^2 g^{-1}}$ and assume that free-free process
dominates the emission and absorbtion. Above and to the right of
each contour, i.e. the larger and more massive progenitors,
$\eta_0<1$ and the radiation is at thermal equilibrium at the shock
breakout time. Progenitors to the bottom and left of the contour
(less massive and more compact) are out of thermal equilibrium
during the breakout, and typically thermal equilibrium is gained
only after the transition to the spherical phase. Typical
$M_*$-$R_*$ of RSG, BSG, WR and WD progenitors are marked as well.
RSGs (n=1.5) are in thermal equilibrium during breakout of typical
SNe and are expected to fall out of equilibrium only if the SN
kinetic energy is $\gtrsim 5 \cdot 10^{51}$ erg. BSGs (n=3) are on
the borderline and larger BSGs are expected to be in thermal
equilibrium during breakout while less massive and compact BSGs
don't. WRs (n=3) and WDs are out of thermal equilibrium in typical
SNe breakout. }\label{fig2}
\end{figure}

\subsubsection{The effect of bound-free absorbtion and emission}\label{SEC BoundFree}
In our calculations above we assumed that free-free is the dominant
emission and absorbtion process. This is the case in low metalicity
progenitor envelopes ($\lesssim 0.1$ solar) but not in high
metalicity ones ($\gtrsim $ solar). When bound-free dominates, the
absorbtion coefficient, $\kappa_{bf}$, is a non-trivial function of
the wavelength, temperature and density. Nevertheless, a rough
estimate of the opacity at $\nu  \approx kT/h$ show that the general
dependence on $T$ and $\rho$ is similar to that of free-free, i.e,
$\kappa_{bf} \propto \kappa_{ff} \propto \rho T^{-3.5} $
\citep[e.g.,][]{Schwarzschild58}. Therefore, when bound-free
dominates, thermalization is achieved faster (and kept for longer)
by a factor $\kappa_{bf}/\kappa_{ff}$ implying that in each shell
$\eta$ is reduced by this factor. Thus, all the equations derived
above are valid after the definition of $\eta$ (eq. \ref{EQ eta
Def}) and the value of $\eta_0$ (eq. \ref{EQ etat_i}) are divided by
$\kappa_{bf}/\kappa_{ff}$ (ignoring the effect of the logarithmic
factor $\xi$).

The correction to the observed temperature in case that the
radiation is in thermal equilibrium ($\etah<1$) is small. The reason
is that the dependence of $T_{obs}$ on $\eta_0$ is very weak
(roughly as $\eta_0^{0.14}$), while in typical envelopes, which are
dominated by Hydrogen or Helium, $\kappa_{bf}/\kappa_{ff} \lesssim
10$. This factor, however, is important in determining whether the
radiation is in thermal equilibrium during the breakout and in
calculating the temperature when it is not. Note that when the
temperature is very high, $\gtrsim 10$ keV, the metals are fully
ionized and bound-free process can be ignored regardless of
metalicity.

\section{The initial pulse}\label{SEC initial pulse}

In previous sections we calculated the luminosity and temperature of
the expanding gas as a function of time. The initial timescale is
$t_0$, and both the luminosity and the typical photons energy evolve
as power-laws of time for $t>t_0$. However, several effects can
smear the observed flux at early times. Differences in the light
travel time to the observer, asphericity of the explosion
\citep[e.g.,][]{Couch10}, or diffusion of the radiation through an
optically thick surrounding such as thick wind blown by the
progenitor before the explosion \citep[e.g.,][]{Li07}. Below we
discuss how light travel time and stellar wind shape the luminosity
and spectrum of the initial pulse.

\subsection{Luminosity}

\subsubsection{Light travel time of spherical explosion}
Light travel time in a spherical explosion significantly affects the
light curve and spectrum only at $t<R_*/c$ and it would be important
only if $R_*/c>t_0$.  The light crossing time is always shorter than
$t_s$ ($R_*/c<R_*/v_0=t_s$), and therefore, its effects are
important only during the planar phase. If  light travel time shapes
the initial pulse then for $R_*/c>t_0$, the observed flux rises over
a timescale $\tdi$ and remains roughly constant over a duration
$R_*/c$. The observed luminosity during the initial pulse is $\sim
L_0 \tdi c/R_* < L_0$. After this plateau, at $t>R_*/c$, the
luminosity starts decreasing as $t^{-4/3}$ and is given again by
equation (\ref{EQ Luminosity}) as light travel time matters no more.
An example of such light curve can be seen in figure \ref{fig Xray
lightcurve} where the WR X-ray luminosity is bolometric until $t_s =
90$ s.

Note, that when light travel time shapes the initial pulse, then
both $R_*$ and $t_0$ can be directly measured. The former is given
by the duration of the initial pulse and the latter by its rise
time. For more extended RSG progenitors, where $\tdi$ may be
comparable to, or exceed, $R_*$ (see appendix A) both the rise time
and duration of the initial pulse are given by $t_0$ and the
progenitor's radius is not directly measurable.

The discussion above is focused on the bolometric luminosity. For
observations at frequencies lower than the initial typical energy
given by $T_{obs,0}$, the rise time will be longer than $\tdi$
(assuming $\tdi<R_*/c$). Depending on thermal equilibrium it is
either $R_*/c$ or the time that the typical temperature falls into
the the observed frequency window (see e.g., figures \ref{fig
optical lightcurve} and \ref{fig Fuv lightcurve}).

\subsubsection{Winds}
WR progenitors are surrounded by the thick stellar wind ejected
during the WR phase.  The optical depth of a wind is gained mostly
close to its source, between $R_*$ and $2R_*$. Typical WR winds, are
mildly optically thick, with optical depth that can be as high as
$\tau_w \sim 1-10$ once the wind is fully ionized by the precursor
of the breakout emission \citep{Li07}. If the wind is very thick,
$\tau_w \gg c/v_{sh}$ the radiation dominated shock would propagate
in the wind as well and it should be treated in a similar way to our
discussion in the previous sections. If $1\lesssim \tau_w \lesssim
c/v_{sh}$ then photons diffuse through the wind without generating a
radiative dominated shock. The energy output of the shock breakout
is not affected but the arrival time of the photons to the observer
is smeared over $\tau_w R_*/c$. The pulse rise and decay times are
both $\tau_w R_*/c$. The information about $\tdi$ is lost and so is
the ability to measure $R_*$ directly.

\subsection{Spectrum}

\subsubsection{Light travel time}
If light travel time shapes the initial pulse, then at first, $t
\sim \tdi$, the spectrum is dominated by the emission from the shock
front which is propagating in a decreasing optical depth during the
breakout. The typical observed photon frequency is therefore
$T_{obs,0}$. As time evolves, $\tdi<t<R_*/c$, the spectrum broadens
to include lower frequencies as well. During this time high
frequency breakout photons, $T_{obs,0}$, continues to arrive from
areas with longer travel time while lower frequency photons from the
expanding gas are arriving from areas with shorter travel time.
Ignoring light travel time effects we derived $T_{obs}=T_{obs,0}
(t/\tdi)^{-\alpha}$, ($1/3<\alpha<2/3$), while $L = L_0
(t/\tdi)^{-4/3}$ (see equations (\ref{EQ Tobs planar}) and (\ref{EQ
Luminosity})). Thus, the spectrum of the initial pulse broadens in
time to form a power-law,
\begin{equation}\label{EQ Fnu initial pulse}
    F_\nu \propto \nu^{\frac{1}{3\alpha}-1},
\end{equation}
over a frequency range that grows with time. Its upper frequency
corresponds to the initial temperature $T_{obs,0}$ while the lower
end of the power-law corresponds to the current (non delayed)
temperature $T_{obs,0}(t/t_0)^{-\alpha}$. The integrated spectrum of
the initial pulse will show this power-law over a frequency range
$kT_{obs,0}(c\tdi/R_*)^\alpha<h\nu<kT_{obs,0}$.

\subsubsection{Wind}
In the case of a mildly opaque wind ($1\lesssim \tau_w \lesssim
c/v$) photons spend a time $\tau R_*/c$ diffusing through the wind,
thereby erasing all the temporal details on shorter time scales. As
a result the observed spectrum is given by  $F_\nu \propto
\nu^{\frac{1}{3\alpha}-1}$ over the frequency range
$kT_{obs,0}(c\tdi/\tau_w R_*)^\alpha<h\nu<kT_{obs,0}$ right from the
beginning.

 Compton and inverse Compton scattering in the wind may also
modify the photon's energy. However, the number of collisions per
photon is $\tau_w^2$, and the number of collisions  needed to
significantly change a photon energy is $m_ec^2/kT$.  Since the
winds we are dealing with have moderate optical depth and our
temperature calculations are applicable to cases where $T \lesssim
50$ keV, scattering within the wind cannot make a significant change
to the energy of such photons. Therefore this effect is not very
important in the temperature range that we can calculate.

\section{Early SNe light curves from various progenitors}\label{SEC lightcurve}

Below we present early light curves (luminosity and spectrum) for
different SN progenitors. We consider different progenitors of
core-collapse SN: red supergiant (RSG), blue supergiant (BSG) and
Wolf-Rayet (WR) stars. We also discuss the effect of deviation from
thermal equilibrium on the signal that follows the shock breakout of
type Ia SN presented in \cite{Piro09}. The light curves are derived
according to the results presented in previous sections, assuming
that free-free is the dominant emission and absorbtion process, and
the properties of the \bo shell given in appendix A.

\begin{figure}[h!]
\centerline{\includegraphics[width=9cm]{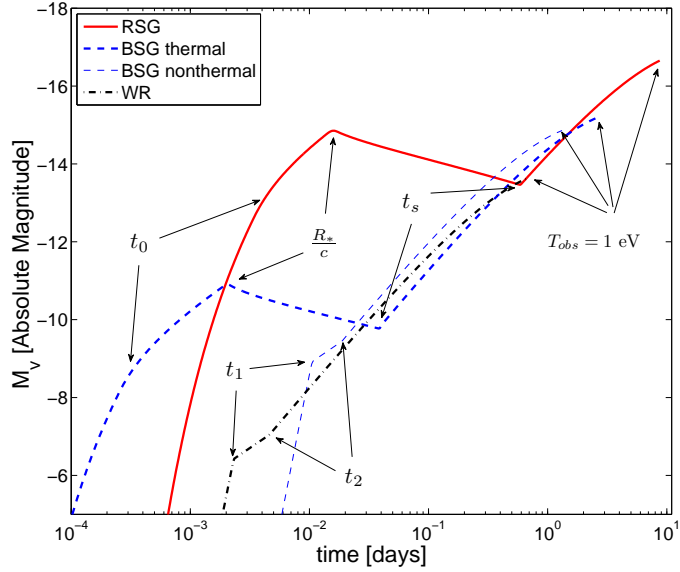}} \caption{The
optical light curve (in absolute V magnitude) following the shock
breakout from RSG ({\it solid line}), BSG in thermal equilibrium
({\it thick dashed line}) and out of thermal equilibrium during the
breakout ({\it thin dashed line}) and WR ({\it dash-doted line}). In
all cases the explosion energy is $10^{51}$ erg. The progenitors
radii are $500 {\rm~ R_\odot}$ (RSG), $70 {\rm~ R_\odot}$ (BSG
thermal), $20 {\rm~ R_\odot}$ (BSG non-thermal) and $5 {\rm~
R_\odot}$ (WR). The progenitors masses are $15 {\rm~ M_\odot}$ in
all cases except for the BSG thermal, where it is $25 {\rm~
M_\odot}$. The explosion is assumed to be spherical (light
travel-time effects are included) and the pre-explosion stellar wind
to be transparent. The source luminosity (before light travel time
effects are included) at $t<t_0$ is approximated as $L_0\exp
[1-\frac{t_0}{t}]$ (this is the luminosity of a radiation that leaks
from the center of a static slab with a diffusion time $t_0$). The
two thermal breakouts (RSG and BSG thermal) show a rising flux over
a duration of $R_*/c$. The rise does not stop at $t_0$ since the
optical band is below $T_{obs,0}$ (the optical is not the bolometric
luminosity at early time). At $R_*/c$ the light curve start decaying
until $t_s$, when the flux start rising again during the spherical
phase up to the time when $T_{obs}$ drops to the observed frequency
(not seen here because we cut the light curves at $T_{obs}=10^4$ K,
see below). In the nonthermal breakouts the flux is strongly
suppressed before $t_s$ since the temperature is very high (much
higher than the optical). Then it rises very sharply up to $t_1$,
when it falls into thermal equilibrium. At $t_2$ the flux joins the
evolution of the thermal breakouts. Thus in the nonthermal breakouts
the flux is rising continuously until the time that $T_{obs}$ drops
to the observed frequency. We present light curves up to the point
that $T_{obs}=10^4$ K, since this is roughly the point where
recombination, which is neglected in our model, becomes important.}
\label{fig optical lightcurve}
\end{figure}

\begin{figure}[h!]
\centerline{\includegraphics[width=9cm]{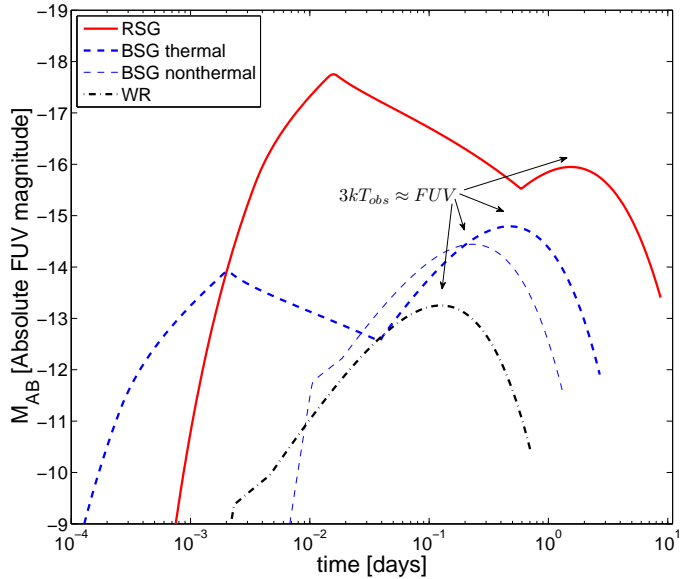}} \caption{The FUV
($\nu_{_{FUV}}=2\cdot 10^{15}$ Hz) light curve in absolute AB
magnitude, following the shock breakout from RSG ({\it solid line}),
BSG in thermal equilibrium ({\it thick dashed line}) and out of
thermal equilibrium during the breakout ({\it thin dashed line}) and
WR ({\it dash-doted line}). The parameters and assumption are the
same as in figure \ref{fig optical lightcurve}. The light curves
evolution is similar to optical ones, with the difference that the
second peak in the flux is observed earlier, once $T_{obs}$ get into
the FUV range.} \label{fig Fuv lightcurve}
\end{figure}

\begin{figure}[h!]
\centerline{\includegraphics[width=9cm]{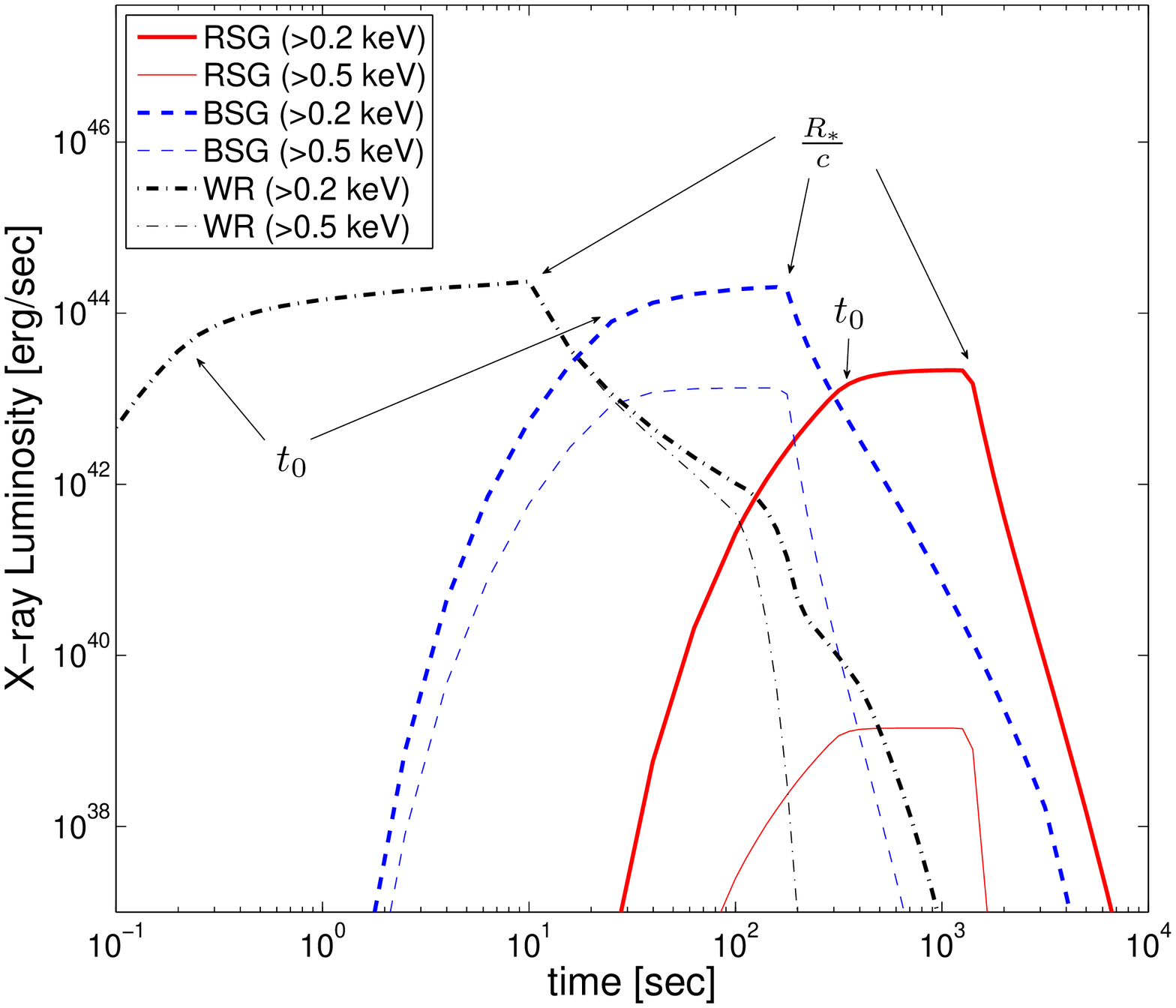}} \caption{The total
luminosity at frequencies $> 0.2$ keV ({\it thick lines}) and $>0.5$
keV ({\it this lines}) following the shock breakout from RSG ({\it
solid lines}), BSG  in thermal equilibrium ({\it dashed lines}) and
WR ({\it dash-doted lines}) stars. The parameters and assumption are
the same as in figure \ref{fig optical lightcurve}. The depicted
luminosity is the bolometric one, at $t<R_*/c$ in the WR and BSG
cases, showing the light curve of the initial pulse in the case of a
spherical explosion. The luminosity rises until $t_0$ when it
becomes almost constant. The decay of the initial pulse starts at
$R_*/c$.} \label{fig Xray lightcurve}
\end{figure}

\subsection{Red supergiant}
RSG is the progenitor of several members of the type II SN family.
It has a convective envelope and its structure can be approximated
using $n=1.5$. We assume a hydrogen envelope with cosmic abundances
so the scattering cross-section per unit of mass is $\kappa = 0.34
{\rm ~cm^2 g^{-1}}$ (the dependence on $\kappa $ is weak). We
consider a typical radius of $500 R_\odot$ (a light crossing time of
about $20 {\rm ~min}$) and a typical mass of $M_*=15 M_\odot$.
Following the initial pulse the luminosity evolves as:
\begin{equation}\label{EQ L_RSG}
    L_{RSG}=\left\{ \begin{array}{lr}
              \,\quad10^{44}  {\rm ~erg/s~} M_{15}^{-0.37} R_{500}^{2.46} E_{51}^{0.3} t_{hr}^{-4/3} & t<t_s \\
              3 \cdot 10^{42}   {\rm ~erg/s~}  M_{15}^{-0.87} R_{500} E_{51}^{0.96} t_{d}^{-0.17} & t>t_s
            \end{array} \right.
\end{equation}
where $R_x=R_*/xR_\odot$, $M_{x}=M_*/x M_\odot$, $E_{x}=E/10^{x}
{\rm ~erg}$ and $t_{hr}$ [$t_d$] is time in units of hours [days].
$E$ here is the explosion energy, not to be confused with the
internal energy in the expanding shells. The transition between the
planar and spherical phases takes place around
\begin{equation}\label{EQ t0_RSG}
    t_s=14 {\rm ~hr~} M_{15}^{0.43} R_{500}^{1.26} E_{51}^{-0.56}
\end{equation}
The value of the thermal coupling parameter at the breakout is
$\eta_0 = 0.06~ M_{15}^{-1.72} R_{500}^{-0.76} E_{51}^{2.16}$.
Therefore, the observed temperature is determined at the outermost
shell which is in thermal equilibrium\footnote{In extreme cases
(e.g., very energetic explosion with $E_{51}>5$), $\eta_0$ may be
larger than unity and the light curve would show a similar evolution
to the one discussed in the context of BSG out of thermal
equilibrium (see below).} and it is given by eqs. \ref{EQ Tobs
planar} and \ref{EQ Tobs spherical small etat}:
\begin{equation}\label{EQ T_RSG}
    T_{RSG}=\left\{ \begin{array}{lr}
              10 {\rm ~eV~} M_{15}^{-0.22} R_{500}^{0.12} E_{51}^{0.23} t_{hr}^{-0.36} & t<t_s \\
              \,\,\, 3 {\rm ~eV~} M_{15}^{-0.13} R_{500}^{0.38} E_{51}^{0.11} t_{d}^{-0.56} & t_s<t
            \end{array} \right.
\end{equation}

The ratio of the diffusion time of the breakout layer and the star
light-crossing time is $c \tdi/R_* = 0.25 M_{15}^{0.21}
R_{500}^{1.16} E_{51}^{-0.79}$, implying that if the explosion is
spherical the initial pulse has a rather well defined observed
temperature of $T_{0,RSG} \approx 25 {\rm ~eV~} M_{15}^{-0.3}
R_{500}^{-0.65} E_{51}^{0.5}$ and its rise time and duration are not
too different.

Typical optical and FUV light curves are depicted in figures
\ref{fig optical lightcurve} and \ref{fig Fuv lightcurve}.  The
initial rise-time in both bands is $R_*/c$ (assuming a spherical
explosion). Note that unlike the bolometric luminosity, here there
is no plateau between $t_0$ and $R_*/c$ since the temperature is
above the observed band at this time. Thus $t_0$ cannot be easily
recovered from optical/UV light curves. The temperature remains
above both the optical and the FUV bands during the planar phase and
both light curves decay slowly. During the spherical phase the
luminosity drops more slowly while $T_{obs}$ drops faster, as a
result a break in the light curve is observed at $t \approx t_s$ and
the optical flux starts rising. It peaks once $T_{obs}$ drops into
the observed frequency range. The optical peak takes place after
about two weeks. At that point the temperature is low enough so
recombination, which we neglected, begins to be important. Therefore
we terminate the light curve in the figures at earlier time when
$T_{obs}=1$~eV.

The X-ray light curve is depicted in figure \ref{fig Xray
lightcurve}. $T_{RSG}(t_0)$ is below the X-ray range and therefore
only the initial pulse may be observed in soft X-rays. We assume
here that the spectrum of the photons in the breakout layer is
thermal, which implies that the initial X-ray pulse is very soft as
the X-ray probes the exponential tail of the spectrum. Nevertheless,
the energy in the soft X-ray flash from an RSG breakout is $\sim 3
\cdot 10^{46}$ erg, which may be detectable to a substantial
distance (possibly larger than that of a WR breakout X-ray flash).

Note that our analysis is appropriate only for RSGs with a density
profile that drops sharply with $d$ near the edge, so the shock is
accelerating before breakout. Our analysis also assumes that the
width of the breakout shell is much smaller than the stellar radius
($d_0 \ll R_*$). Stellar structure models such as those calculated
by \cite{Gezari08}, using the KEPLER code \citep{Weaver78} and 3D
hydrodynamics code \citep{Freytag02}, are compatible with these
assumptions. Our model assumptions, however, are incompatible with
density profiles such as the one used by \cite{Schawinski08}, where
the density of the envelope is very low ($\sim 10^{-11} {\rm
gr/cm^3}$ at $0.5 R_*$) and $d_0 \sim R_*$.

\subsection{Blue supergiant}
BSG is also a progenitor of member(s) in the type II SN Family. It
has a radiative envelope and is well approximated using $n=3$. We
assume a hydrogen envelope with $\kappa$ similar to RSG. The typical
radius is $50 R_\odot$ giving $R_*/c \approx 2$ min. Following the
initial pulse the luminosity evolves as:
\begin{equation}\label{EQ L_BSG}
    L_{BSG}=\left\{ \begin{array}{lr}
              2.5 \cdot 10^{44} {\rm ~erg/s~} M_{15}^{-0.33} R_{50}^{2.3} E_{51}^{0.34} t_{min}^{-4/3} & t<t_s \\
              \,\,\,\, 2 \cdot 10^{42} {\rm ~erg/s~} M_{15}^{-0.73} R_{50} E_{51}^{0.91} t_{hr}^{-0.35} & t>t_s
            \end{array} \right.
\end{equation}
where $t_{min}$ is time in minutes. The transition between the
planar and spherical phases takes place around
\begin{equation}\label{EQ t0_BSG}
    t_s=0.5 {\rm ~hr~} M_{15}^{0.41} R_{50}^{1.33} E_{51}^{-0.58}
\end{equation}
The value of the thermal coupling parameter at the breakout is
$\eta_0 \approx 2~ M_{15}^{-1.63} R_{50}^{-1.1} E_{51}^{2.24}$.
Therefore, BSG progenitors are on the thermal coupling borderline.
Thus, we expect both types of BSG shock breakouts, where less
energetic explosions of more extended and massive BSGs will be in
thermal equilibrium, while more energetic explosions of compact less
massive BSGs will be out of thermal equilibrium.  For example an
$E=10^{51}$ erg  explosion of a $R_*=70 R_\odot$ and $M_*=25
M_\odot$ progenitor has $\eta_0 \approx 0.5$, and is therefore
predicted to be in thermal equilibrium while if $R_*=20 R_\odot$ and
$M_*=10 M_\odot$ then $\eta_0 \approx 10$ and the breakout is
predicted to be out of thermal equilibrium. Finally, in high
metalicity envelopes bound-free emission dominates over free-free
emission and $\eta_0$ is reduced (section \ref{SEC BoundFree}).
Therefore we provide here the temperature evolution for both thermal
and non-thermal cases.

If the radiation is in thermal equilibrium at breakout time, $\eta_0
< 1$ (equations. \ref{EQ Tobs planar} and \ref{EQ Tobs spherical
small etat}) the observed temperature is:
\begin{equation}\label{EQ T_BSG}
    T_{BSG}( \eta_0<1)= \left\{  \begin{array}{lr}
              50 {\rm ~eV~} M_{25}^{-0.19} R_{70}^{0.06} E_{51}^{0.22} t_{min}^{-16/45} & t<t_s \\
              10 {\rm ~eV~} M_{25}^{-0.11} R_{70}^{0.38} E_{51}^{0.11} t_{hr}^{-0.61} & t_s<t
            \end{array} \right. .
\end{equation}

In case that the radiation is out of thermal equilibrium at breakout
time, $\eta_0 > 1$, we use eqs. (\ref{EQ Tobs planar}), (\ref{EQ
Tobs spherical t2}) and (\ref{EQ Tobs spherical small etat}) to
find:
\begin{equation}\label{EQ T_BSG no eq}
   \begin{array}{l}
    T_{BSG}(\eta_0>1) \\
    =\left\{  \begin{array}{lr}
              150 {\rm ~eV~} M_{10}^{-1} R_{20}^{-0.1} E_{51}^{1.2} t_{min}^{-0.39} & t<t_s \\
              70 {\rm ~eV~} M_{15}^{-0.9} R_{20}^{-0.7} E_{51}^{1.1} \left(\frac{t}{t_s}\right)^{-2.1} & t_s<t<t_1\\
              15 {\rm ~eV~} M_{10}^{0.05} R_{20}^{0.25} E_{51}^{-0.1} \left(\frac{t}{15{\rm ~min}}\right)^{-0.4} & t_1<t<t_2\\
              7 {\rm ~eV~} M_{10}^{-0.11} R_{20}^{0.38} E_{51}^{0.11} t_{hr}^{-0.61} & t_2<t
            \end{array} \right.   \end{array} ,
\end{equation}
where
\begin{equation}\label{EQ t2_BSG}
    t_1=13 {\rm ~min~} M_{10}^{-0.24} R_{20}^{0.94} E_{51}^{0.29}
\end{equation}
and
\begin{equation}\label{EQ t2_BSG}
    t_2=20 {\rm ~min~} M_{10}^{-0.77} R_{20}^{0.62} E_{51}^{0.99}
\end{equation}
as given by equations (\ref{EQ t1}) and (\ref{EQ t2}). The
temperature at $t<t_1$ (when the observed radiation is out of
thermal equilibrium) depends on the value of the logarithmic
Comptonization factor, $\xi$, which does not evolve as a power-law.
Thus, the temperature decay during the planar phase and early
spherical phase ($t<t_1$) is not an exact power-law. Here we provide
a power-law approximation, using an average temporal indices
$\alpha_{_{0s}}$ and $\alpha_{_{s1}}$ ($T_{BSG}(t<t_s)\propto
t^{-\alpha_{_{0s}}}$, $T_{BSG}(t_s<t<t_1)\propto
t^{-\alpha_{_{s1}}}$). The values we use here, $\alpha_{_{0s}} =
0.39$ and $\alpha_{_{s1}}=2.1$, are for the specific choice of the
canonical explosion parameters, obtained by solving equation \ref{EQ
Tobs planar} numerically. As discussed above
$1/3<\alpha{_{0s}}<2/3$, and is increasing when the breakout is
farther from equilibrium. If for example we take an explosion of
$E_{51}=5$ (keeping $M_{10}=1$ and $R_{20}=1$) we find an average
value of $\alpha_{_{0s}}=0.48$ and $\alpha_{_{s1}}=2$. Due to the
dependence on $\xi$ the power-law dependence of $T_{BSG}(t<t_1)$ on
the explosion parameters ($M_{10}$, $R_{20}$ and $E_{51}$) is also
approximated and calculated numerically. The power law indices of
$M_{10}$, $R_{20}$ and $E_{51}$ that we provide are accurate to
within $\pm 0.5$ in the range $M_{10}=0.3-3$, $R_{20}=0.75-5$ and
$E_{51}=1-2$.

The ratio of the diffusion time of the breakout layer and the star
light crossing time is $c \tdi/R_* = 0.09   M_{15}^{0.27}
R_{50}^{0.91} E_{51}^{-0.73}$. Thus, if $\eta_0 < 1$ the observed
temperature in the initial pulse smoothed between $T_{0,BSG}$ and
$0.5T_{0,BSG}$, while for $\eta_0 > 1$ it is smoothed between
$T_{0,BSG}$ and $0.2T_{0,BSG}$, where
\begin{equation}
    T_{0,BSG}\approx \left\{\begin{array}{lc}
                         80  {\rm ~eV~} M_{25}^{-0.28} R_{70}^{-0.62} E_{51}^{0.48} & \eta_0 < 1 \\
                         700 {\rm ~eV~} M_{10}^{-1.2} R_{20}^{-1.1} E_{51}^{1.7} & \eta_0 > 1
                       \end{array}\right.
\end{equation}
Similarly to equation \ref{EQ T_BSG}, The power law indices of
$M_{10}$, $R_{20}$ and $E_{51}$ in case that $ \eta_0 > 1$ are
approximated and calculated numerically, with the same accuracy.

Typical optical and FUV light curves are depicted in figures
\ref{fig optical lightcurve} and \ref{fig Fuv lightcurve}. The light
curves in case that thermal equilibrium is assumed are similar to
the RSG case (with a different values of $R_*/c$ and $t_s$) and they
show an initial rise over a duration $R_*/c$, followed by a slow
decay up to $t_s$. In the non-thermal case the emission is strongly
suppressed (as $T_{obs} \gg T_{BB}$) and the optical/FUV flux is
rising sharply up to $t_1$ (slightly after $t_s$) where $T_{obs} =
T_{BB}$. As expected once $T_{obs} = T_{BB}$  the two light curves,
thermal and non-thermal, show a similar evolution. Note that the
temperature at a given observer time during this phase depends
mostly on the progenitor radius and therefore the peak in the light
curve, which is observed once $T_{obs}$ drop to the observed
frequency, is observed at earlier time for more compact progenitor.

The X-ray light curve when thermal equilibrium is assumed at all
time, is depicted in figure \ref{fig Xray lightcurve}. $T_{0,BSG}$
is $\sim 0.1$ keV and the X-ray luminosity is comparable to the
bolometric luminosity. If the breakout emission is out of thermal
equilibrium $T_{0,BSG}$ is higher and the observed x-ray spectrum is
harder. But in both cases most of the breakout luminosity, $\sim 3
\cdot 10^{46}$ erg, is expected to fall within the X-ray range.
Making the X-ray flash from BSG breakouts the easiest to observe
among the different progenitor types.

\subsection{Wolf-Rayet}
The core collapse of a WR star is most likely the onset of a type
Ib/c SN. At WR stage the star radius is only several solar radii
($R_*/c \sim 10 {\rm s}$) and its mass is 10-80 $M_\odot$. The
envelope is radiative (n=3)
 and has no hydrogen ($\kappa = 0.2 {\rm~ cm^2
g^{-1}}$). In cases where the WR wind is very thick it may be dense
enough to provide significant opacity and play a role during the
short planar phase (see section \ref{SEC initial pulse}). Note that
this effect vanishes during the spherical phase, where $\tauh$
increases while the wind opacity decreases. Neglecting possible wind
opacity effects the luminosity following the initial pulse evolves
as:
\begin{equation}\label{EQ L_BSG}
    L_{WR}=\left\{ \begin{array}{lr}
              \,\,\,\, 2 \cdot 10^{42} {\rm ~erg/s~} M_{15}^{-0.33} R_{5}^{2.3} E_{51}^{0.34} t_{min}^{-4/3} & t<t_s \\
              3.5 \cdot 10^{41} {\rm ~erg/s~} M_{15}^{-0.73} R_{5} E_{51}^{0.91} t_{hr}^{-0.35} & t>t_s
            \end{array} \right.
\end{equation}
where
\begin{equation}\label{EQ t0_WR}
    t_s=90 {\rm ~s~} M_{15}^{0.41} R_{5}^{1.33} E_{51}^{-0.58}
\end{equation}
The value of the thermal coupling parameter at the breakout is
$\eta_0 = 24~ M_{15}^{-1.63} R_{5}^{-1.1} E_{51}^{2.24}$, implying
that initially the radiation is not in thermal equilibrium with the
gas and the observed temperature is given by eqs. (\ref{EQ Tobs
planar}), (\ref{EQ Tobs spherical t2}) and (\ref{EQ Tobs spherical
small etat}):
\begin{equation}\label{EQ T_WR}
    T_{WR}=\left\{ \begin{array}{lr}
              1 {\rm ~keV~} M_{15}^{-1.5} R_{5}^{-0.2} E_{51}^{1.4} t^{-0.4} & t<t_s \\
              140 {\rm ~eV~} M_{15}^{-1.2} R_{5}^{0.9} E_{51}^{1.7} \left(\frac{t}{t_s}\right)^{-2.2} & t_s<t<t_1\\
              40 {\rm ~eV~~} M_{15}^{0.05} R_{5}^{0.25} E_{51}^{-0.1} t_{min}^{-0.4} & t_1<t<t_2\\
              ~5  {\rm ~eV~~} M_{15}^{-0.11} R_{5}^{0.38} E_{51}^{0.11} t_{hr}^{-0.61} & t_2<t
            \end{array} \right.
\end{equation}
where
\begin{equation}\label{EQ t2_WR}
    t_1=200 {\rm ~s~} M_{15}^{-0.24} R_{5}^{0.94} E_{51}^{0.29}
\end{equation}
and
\begin{equation}\label{EQ t2_WR}
    t_2=400 {\rm ~s~} M_{15}^{-0.77} R_{5}^{0.62} E_{51}^{0.99}
\end{equation}
Similarly to the nonthermal BSG case the temperature at $t<t_1$ does
not evolve as a power-law, so the values $\alpha_{_{0s}} = 0.4$ and
$\alpha_{_{s1}}=2.2$, are approximations obtained for the specific
choice of the canonical explosion parameters  (see the discussion
below equation \ref{EQ t2_BSG}). For the same reason the power-law
dependence on the explosion parameters at $t<t_1$ is approximate,
where the power-law indices of $M_{15}$ and $R_{5}$ are accurate to
within $\pm 0.5$ and of $E_{51}$ to within $\pm 0.8$ in the range
$M_{15}=0.2-2$, $R_{5}=0.2-2$ and $E_{51}=1-2$.

The observed temperature at the first few minutes is much higher
than the prediction of a model that assumes thermal equilibrium at
all time, peaking at the breakout temperature:
\begin{equation}\label{EQ T0_WR}
    T_{0,WR} \approx 2 {\rm ~keV~} M_{15}^{-1.7} R_{5}^{-1.5} E_{51}^{1.8}
\end{equation}
Similarly to $T_{obs}(t<t_s)$ the dependence of $T_{obs,0}$ on the
explosion parameters is approximated and calculated numerically with
a similar accuracy to the indices provided for $T_{obs}(t<t_s)$.
Finally, Note that our model, which neglects relativistic effect, is
not applicable at $T>50$ keV. Thus whenever the model predicts
higher temperature it overestimates the true temperature which does
not exceed $\sim 200$ keV.

In a typical WR star $c \tdi/R_* = 0.014 M_{15}^{0.27} R_{5}^{0.91}
E_{51}^{-0.73}$. Therefore the initial pulse is always composed of a
range of temperatures (regardless of the wind opacity). The
integrated spectrum shows a non-thermal power-law, $F_\nu \propto
\nu^{-\beta}$ where $0<\beta<0.5$, over more than one order of
magnitude at $h\nu<k T_{0,WR}$.

Typical optical and FUV light curves are depicted in figures
\ref{fig optical lightcurve} and \ref{fig Fuv lightcurve}. The flux
in these bands is very faint at $t<t_s$ (as long as the emitted
radiation is out of thermal equilibrium) and it rises continuously
until $T_{obs}$ drops into the observed band. Figure \ref{fig Xray
lightcurve} depict the observed flux in X-rays and soft gamma-rays.
This energy range contains almost all of the breakout luminosity.
The radiation from less energetic breakouts from larger WR
progenitors will all be in the range of X-ray  detectors (0.2-10
keV), while more energetic breakouts from compact WR progenitors
will be in the range of soft gamma-ray detectors($>10$ keV). Thus,
we expect that at least some WR breakouts can be detected as soft
gamma-ray bursts by satellites such as Swift up to a distance of
$\sim 10$ Mpc.

\subsection{White dwarf}\label{SEC WD}
The thermo-nuclear explosion of a white dwarf near the Chandrasekhar
mass is most likely the origin of a type Ia SN. The shock breakout
in this case was carefully explored recently by \cite{Piro09}. This
paper assumes a thermal equilibrium at all time, while pointing out
that this assumption may be violated at early times. Here we repeat
the main results of \cite{Piro09}, highlighting the effects of the
deviation from thermal equilibrium at early times.

The shock velocity at breakout may be either mildly relativistic
($v_0/c [1-v_0^2/c^2]^{-1/2} \approx 1$) or relativistic
($[1-v_0^2/c^2]^{-1/2} \gg 1$) \citep{Tan01}. Here we consider only
cases where the breakout is mildly relativistic and therefore a
Newtonian approximation of the hydrodynamics is reasonable for the
breakout properties. in such case $t_s \approx R_*/c$ and the energy
in the initial pulse is $E_0 \sim 10^{40}-10^{41}$ erg over a
duration $R_*/c \approx t_s \approx 10-20$ ms. The initial diffusion
time is much shorter ($\tdi \sim {\rm \mu sec}$), implying that if
asphericity does not play a major role, the bolometric luminosity
rises practically instantaneously. The temporal evolution of the
luminosity at $t>t_s$ is independent of the assumption of thermal
equilibrium and is given in \cite{Piro09}. The normalizations that
we provide here, based on the simple assumptions described in
appendix A (which ignores the phase during which the nuclear burning
provides energy to the shock), are comparable to the more accurate
calculation of \cite{Piro09}:
\begin{equation}\label{EQ L_WD}
   \begin{array}{l}
     L_{WD} \approx \\
     \left\{ \begin{array}{l}
              5 \cdot 10^{40} {\rm \frac{erg}{s}} M_{1.4}^{-0.73} \frac{R}{5\cdot 10^8 {\rm ~cm}} E_{51}^{0.91}
              t^{-0.35}~~~~~~~~~~~~;~
              t_s<t<t_{deg}\\
              3 \cdot 10^{39} {\rm \frac{erg}{s}} M_{1.4}^{-0.73} \frac{R}{5\cdot 10^8 {\rm ~cm}} E_{51}^{0.91} \left(\frac{t_{deg}}{700 s}\right)^{-0.18}
              t_{hr}^{-0.17}~;~       t_{deg}<t
            \end{array} \right. ,
   \end{array}
\end{equation}
where $t_{deg}$ is the time in which $\mh$ makes the transition from
mass that was not degenerate before the explosion ($n=3$) to matter
that was degenerate (n=1.5). In the model discussed by \cite{Piro09}
$t_{deg} \approx 700$ s.

The breakout temperature, assuming thermal equilibrium is $T_{BB,0}
\approx 10$ keV, but $\eta_0 \sim 10^5$ implying first that the
system is far from thermal equilibrium and second that relativistic
effects, such as pair production, play a major role. Therefore we
cannot determine the exact observed spectrum at early time, but pair
production limits the temperature at this regime to be
\citep{Katz09}:
\begin{equation}\label{EQ Tti WD}
  T_{0,WD} \approx 200 {\rm~ keV}
\end{equation}
This is the temperature in the shocked gas frame, and it is also the
observed temperature if the shock is not relativistic. In case that
the shock is relativistic the observed temperature is higher.
Therefore a breakout from a type Ia SN will produce a short flash of
gamma-rays which is easily detectable within our Galaxy by various
gamma-ray satellites such as IPN satellites and Fermi, and may
possibly be detectable out to the Magellanic Clouds by Swift. At
$t>t_s$ the temperature drops quickly until it gets to $T_{BB}$ at
$t_2 \approx 1$ s. The short flash of $\gamma$-rays may be
classified as a short GRB without the following detection of a SN.
At later times the observed temperature is:
\begin{equation}\label{EQ T WD}
     T_{WD} = 6 {\rm ~eV~} M_{1.4}^{-0.11}
     \left(\frac{R}{5\cdot 10^8 {\rm ~cm}}\right)^{0.38}
     E_{51}^{0.11} t_{min}^{-0.61} \,;\,1\ll t < t_{deg} .
\end{equation}
This equation can be extrapolated to $t > t_{deg}$ using $T_{WD}
\propto t^{-0.56}$. Note that only at $t \gg1 {\rm ~s~}$ the
velocity of $\mh$ is significantly lower than $c$ (e.g., $\vh(1 s)
\approx 0.5c$), limiting the applicability of our temperature
evolution calculation to $t \gg1$. At this point the emitted
radiation is already in thermal equilibrium. Yet, our predicted
temperature at late times, when thermal equilibrium holds, does not
agree with this of \cite{Piro09}. They assumed that the temperature
is related to the luminosity and the radius of the luminosity shell,
$T_{obs} = (\Lh/4\pi \sigma \rh^{~2})^{1/4}$. However, as we have
shown, the photons that are emitted from the luminosity shell are
reprocessed within all the external shells that are in equilibrium.
Thus, we use $T_{obs} = T(\eta=1)$ (see the discussion in section
\ref{Sec: spectrum theory}), and predict lower temperatures compared
to \cite{Piro09}. Eq. \ref{EQ T WD} implies that the light curve we
provide is valid only during the first $\sim 20$ min. At later times
the temperature is low enough so recombination can no longer be
neglected.

\section{Comparison to previous works}\label{SEC comparison}
Various aspects of the early supernova light curve were already
addressed by a large number of authors, both analytically and
numerically. The early SNe light curve that we consider here is
composed of three phases: breakout, planar phase and spherical
phase. Almost all analytical calculations focused only on a single
phase. Therefore we first compare our methods and results to
previous analytical calculations phase by phase. Later we compare
our results to those of numerical calculations that include
(implicitly) all the phases and the transitions between them.

\subsection{analytic works}

The method we use to find the \bo shell, i.e., finding
$\tau_0=c/v_0$, is similar to the one used in previous analytical
studies \citep[e.g.,][]{Imshennik88,MatznerMcKee99} and so are the
values of $E_0$ and $L_0$ that we find. Also the temperature of the
\bo shell is similar, but only if the \bo shell is in thermal
equilibrium. The observed \bo temperature depends on whether the
post shock radiation in the \bo shell is in thermal equilibrium or
not. If it is, then the color shell is exterior to the \bo shell but
interior to the layer where $\tau=1$. Therefore, the observed
breakout temperature, $T_{obs,0}$, is lower than $T_{BB,0}$ but
larger than the effective temperature, defined as $T_{eff}\equiv [ L
/ 4\pi \sigma r_{\tau=1}^{~2}]^{1/4}$. This fact was recognized by
several authors \citep[e.g.,][]{Falk78,Ensman92}. They estimate
numerically that during breakout the ratio of the observed
temperature, (denoted sometime as the color temperature $T_c$), to
the effective temperature is  $T_{obs,0}/T_{eff,0} \approx 2$.
Similarly, in our analytic theory we obtain that this ratio is 1.8
for our canonical RSG and 2.1 for our canonical BSG.

We determine if the \bo shell is in thermal equilibrium when the
shock breaks out, by equating the time available for photon
production, $\min\{t,t_d\}$, and the time needed to obtain thermal
equilibrium (we use the same method also at any other shell and
time). This method was used by \cite{Weaver76} to analytically find
departure from thermal equilibrium just behind radiative dominated
shocks, where the available time is roughly the shock crossing time.
During breakout  $t \approx t_d$ in the \bo shell and both are
comparable to the shock crossing time. Therefore, our estimates
should coincide with those of \cite{Weaver76} when the properties of
the shock once it breaks out are taken. Our requirement for
departure from thermal equilibrium at breakout, $\eta_0>1$, is
translated at the densities of interest (the density dependence is
extremely weak) to energy per nucleon $\gtrsim 1$ MeV or a shock
velocity $\gtrsim 15,000$ km/s. This result is similar to the one
obtained by \cite{Weaver76} (his equation 5.15). \cite{Weaver76}
also calculated numerically the temperature in case that the gas
behind the shock velocity is out of equilibrium. \cite{Katz09}
realized the importance of this result to SN shock breakouts and
presented an analytic calculation (their equation 18), which is
similar to our equation \ref{EQ Teta1} (the two equations are almost
identical when the numerical factor in our equation \ref{EQ etat_i}
is taken as 0.4 instead of 0.2 and our $\xi$ is identified as their
$\Lambda_{\rm eff}g_{\rm eff}$). Our analytic calculation is in
excellent agreement with the numerical results of \cite{Weaver76}
over the whole range of relevant densities and velocities before
pair production becomes important.

The only analytical calculation of the planar phase was carried out
by \cite{Piro09}, in the context of SN type Ia  breakout from a
white dwarf. The luminosity we obtain is similar to that of
\cite{Piro09}. However, the observed temperature that we find is
different. \cite{Piro09} assumes thermal equilibrium and
$T_{obs}=\Th_{BB}$. At early time thermal equilibrium is not
achieved and therefore \cite{Piro09} underestimate the observed
temperature. At late time the radiation thermalized further out than
the \lum shell and therefore they overestimate the observed
temperature. A detailed comparison with their work is presented in
\S \ref{SEC WD}.

The spherical phase was explored analytically, using different
methods by a number of authors
\citep{Chevalier92,Chevalier08,Waxman07,Rabinak10,Piro09}. We
calculate the luminosity by identifying the energy source of the
observed radiation as the point where photons can diffuse out over a
dynamical time. This method use the same physical picture as
\cite{Chevalier92} and \cite{Chevalier08}, which carried out a more
detailed calculation by finding self similar solutions for the
diffusion wave that propagates into the ejecta. Their results are
similar to ours. \citealt{Waxman07} and \cite{Rabinak10} use a
different method to calculate the luminosity. They use the equation
$L= 4\pi \sigma r_{\tau=1}^{2} T_{\tau=1}^4$ (which assumes thermal
equilibrium at the $\tau =1$ shell), where $T_{\tau=1}$ is
calculated by pretending that the radiation in the $\tau=1$ shell
has cooled adiabatically since breakout. However, the radiation that
was generated by the shock at that shell is long gone by the time
that $\tau=1$ (it escaped once its opacity satisfied $\tau=c/v$) and
the radiation that arrives to the $\tau=1$ shell from inner shells
is not in thermal equilibrium. This method overestimates the
luminosity by a factor of $\tauh^{~0.08}$-$\tauh^{~0.16}$ for
$n=1.5-3$. For the specific problem and parameters that we consider
here $\tauh=30-70$ during the spherical phase of a core-collapse SN,
translating to an overestimate of the luminosity by up to a factor
of 2.

All previous works assumes that the observed radiation is in thermal
equilibrium during the spherical phase. We find that thermal
equilibrium is always achieved long time after the transition
between the planar and spherical phases.
\cite{Chevalier92,Chevalier08} and \cite{Waxman07} assumes that the
radiation is in thermal equilibrium all the way out to the location
where $\tau=1$. This leads to an underestimate of the observed
temperature by a factor of $\tau(\eta=1)^{0.25}$, which for our
canonical SN parameters is an underestimate by up to a factor of 2
during the spherical phase.

\cite{Rabinak10} explore SN light curves at $T<3$ eV, including the
effect of recombination and taking into account Thomson, free-free
and bound-free opacity. Our calculation, which ignores
recombination, can be compared to their treatment only at $T \gtrsim
1$ eV, where the deviation due to recombination is less prominent.
\cite{Rabinak10} look into the difference between the observed
temperature and effective temperature. They find $T_{obs}/T_{eff}
(1<T<3 {\rm ~eV}) \approx 1.2$, which is slightly lower than the
values that we obtain (1.3-1.5) when bound-free opacity is ignored.

\subsection{numerical works}
Many authors used numerical simulations to study early SN light
curves
\citep[e.g.,][]{Shigeyama88,Woosley88,Ensman92,Blinnikov98,Schawinski08,Tominaga09}.
Two of these works provide bolometric luminosity and observed
temperature at temporal resolution that is high enough for
comparison with our model, starting at the breakout through the
planar and spherical phases. These are \cite{Ensman92} that simulate
BSG explosions and \cite{Tominaga09} that simulate an RSG explosion.

\begin{figure}[h!]
\centerline{\includegraphics[width=9cm]{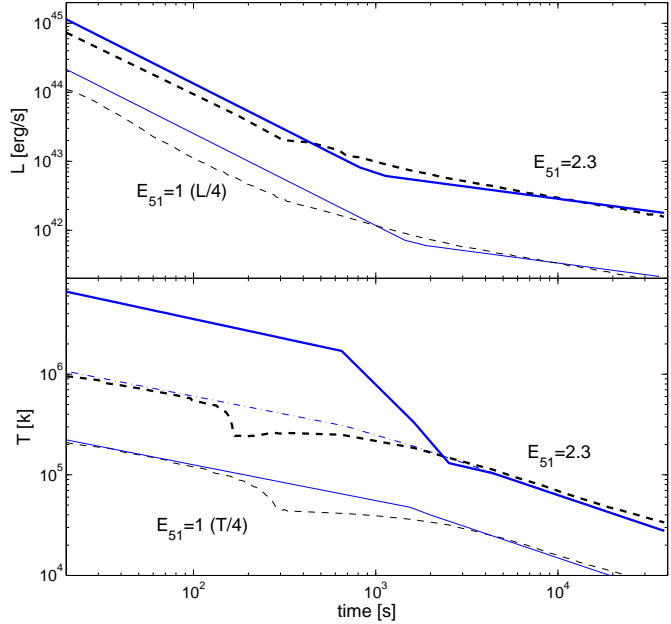}} \caption{Bolometric
luminosity ({\it upper panel}) and observed (color) temperature
({\it lower panel}) of two BSG explosions as a function of time, as
obtained by the numerical simulation of \cite{Ensman92} ({\it dashed
black line}) and by our analytic calculation ({\it solid blue
line}). The progenitor parameters are $M_{15}=16/15$ and
$R_{50}=0.9$. The two explosions energies are $E_{51}=1$ ({\it thin
lines}) and $E_{51}=2.3$ ({\it thick lines}). The luminosity and
temperature of the $E_{51}=1$ explosion are divided by a factor of 4
(both ours and those obtained by \citealt{Ensman92}). The
temperature of the $E_{51}=2.3$ explosion in case that thermal
equilibrium is artificially enforced is also plotted ({\it thin
dashed-dot line}). See text for details.} \label{Fig Ensman92}
\end{figure}

\cite{Ensman92} simulate several  BSG explosions. They provide
detailed luminosity and temperature curves, with no corrections for
light travel time effects, of a BSG progenitor with  $M_*=16~{\rm
M}_\odot$ and $R_*=45~{\rm R}_\odot$  that explodes with energies of
$10^{51}$ erg and $2.3 \cdot 10^{51}$ erg. The temporal resolution
is about $10$ s for the first few hundred seconds, adequate for
comparison with BSG breakout and planar phase evolution. At later
times they provide curves with a temporal resolution of about 0.5
hr. \cite{Ensman92} allow for deviation of the gas temperature from
the  radiation temperatures and identify the color shell (i.e.,
thermalization depth) by the requirement $\sqrt
{3\tau_{abs}\tau}=2/3$. They do however impose
$\epsilon_{rad}=aT_{rad}^4$, where $\epsilon_{rad}$ and $T_{rad}$
are the radiation energy density and temperature respectively,
thereby not allowing for deviation of the radiation from thermal
equilibrium. Figure \ref{Fig Ensman92} shows the bolometric
luminosity and observed temperature as functions of time obtained by
\cite{Ensman92}. These are extracted from their figures 2 \& 11 at
early times and 3 \& 12 at late times. $t=0$ is set such that the
peak of the luminosity is at $t_0$. Figure \ref{Fig Ensman92} also
shows our results of the luminosity and observed temperature. Our
luminosity is calculated using equation \ref{EQ L_BSG} for both
explosion energies. The excellent agreement between \cite{Ensman92}
numerical luminosity calculation and our analytic results is
achieved without any fitting of the normalization or any other
parameter. The results range over three orders of magnitude in time
and luminosity and do not differ more than a factor of 2. The
temporal decay slopes are similar and so are the values during the
planar and spherical phases.

For $E_{51}=1$, the breakout thermal equilibrium is marginal
($\eta_0 \approx 2$) and we plot only the temperature curve expected
if thermal equilibrium is kept at all time (as enforce artificially
by \citealt{Ensman92}) using equation \ref{EQ T_BSG}. For
$E_{51}=2.3$, the breakout radiation is out of thermal equilibrium
($\eta_0 \approx 13$) and therefore we calculate the temperature
using equation \ref{EQ T_BSG no eq}. In order to compare our model
to the results of \cite{Ensman92} we also plot the temperature
predicted in that case if there were thermal equilibrium at all
time, using equation \ref{EQ T_BSG}. The agreement between our model
(when thermal equilibrium is enforced) and the results of
\cite{Ensman92} is very good. It is better than $20\%$ at any time,
except for a period around a few hundred seconds where the
temperatures of \cite{Ensman92}  drop over a short period by a
factor of two. We do not see any physical origin for this behavior,
which may be a numerical artifact. In the more energetic explosion,
$E_{51}=2.3$, our model shows that the assumption of thermal
equilibrium fails at early time and that \cite{Ensman92}
underestimate the breakout and planar phase observed temperature by
an order of magnitude.

\begin{figure}[h!]
\centerline{\includegraphics[width=9cm]{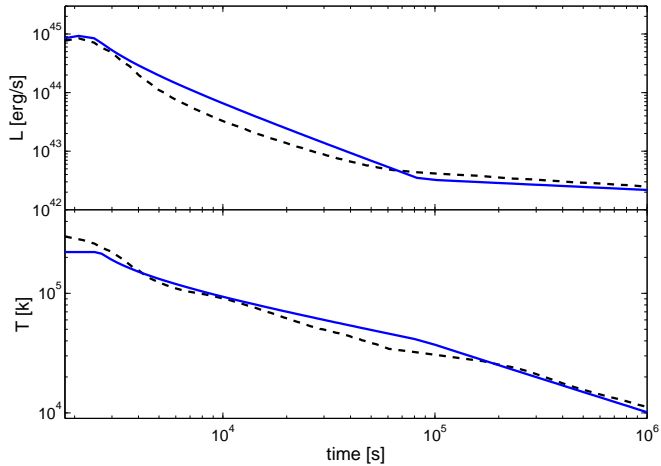}} \caption{Bolometric
luminosity ({\it upper panel}) and observed (color) temperature
({\it lower panel}) of an RSG explosion as function of time as
obtained numerically by \cite{Tominaga09} ({\it dashed black line})
and by our analytic calculation ({\it solid blue line}). The
explosion parameters are $M_{15}=1.2$ and $R_{500}=1.6$ and
$E_{51}=1.2$. The luminosity calculated by our analytic formula
(equation \ref{EQ L_RSG}) is deviled by a constant factor of 1.4.
see text for details} \label{Fig Tominaga09}
\end{figure}

\cite{Tominaga09} simulate a $1.2 \cdot 10^{51}$ erg explosion of an
RSG progenitor with $M_*=18~{\rm M}_\odot$ and $R_*=800~{\rm
R}_\odot$. They provide  bolometric luminosity and observed
temperature curves, where a travel time effects are {\it included},
assuming a spherical explosion. The temporal resolution is about
$0.5$ hr for the first 14 hour and about 1 day at later times. We
extract these curves from their figure 3 and set $t=0$ so the
luminosity peaks at $t=R_*/c = 1866$ s. Figure \ref{Fig Tominaga09}
shows the bolometric luminosity and observed temperature as
functions of time obtained by \cite{Tominaga09} and those obtained
by our calculations. Our luminosity is calculated by using equation
\ref{EQ L_RSG} and smoothing it for photon arrival time effect from
a spherical explosion. The result is then divided by a constant
factor of 1.4 to best fit the luminosity found by \cite{Tominaga09}.
The similarity between the result of the detailed numerical
simulation and our calculation is very good. Again, the luminosity
ranges over three orders of magnitude and it is similar to within a
factor of two at all times (factor of 3 if our luminosity is not
divided by a constant factor of 1.4). The observed temperature is
calculated using equation \ref{EQ T_RSG}. At early times, when light
travel time effects cause the observer  to see a range of
temperatures at any given moment, the maximal of these temperatures
is taken. Our resulting temperature is not multiplied by any
constant factor. The agreement between the numerical and analytic
temperatures is again very good. It is similar at any time to within
30$\%$.

\section{Summary}\label{SEC Summary}
We derive analytic SNe light curves at early times, as long as
recombination and radioactive decay do not play an important role.
These light curves are valid while the observed temperature is above
about $1$ eV and before injection by radioactive decay becomes
important. These conditions hold during the first day after the
explosion of a typical SN. The main advantage of our analysis over
previous ones is the account for the radiation-gas coupling, which
leads to determination of the observed temperature when the
radiation at the color shell is out of thermal equilibrium. It also
corrects previous estimates of the observed temperature, and the
color shell location, when the radiation at the color shell is in
thermal equilibrium. We define a thermal coupling coefficient,
$\eta$, and find that the temperature evolution can follow two very
different tracks, depending on $\eta_0$, i.e., the value of $\eta$
in the breakout shell at the breakout time. When the breakout shell
is out of thermal equilibrium ($\eta_0>1$) the observed temperature
starts high above the value obtained when thermal equilibrium is
assumed and it drops faster than the case that the breakout shell is
in thermal equilibrium. Thermal equilibrium is typically gained
(when $\eta_0>1$) only at early stages of the spherical phase.

We discuss the luminosity and spetral evolution during the initial
pulse and derive early SN light curves for various SN progenitors as
a function of the explosion energy and the progenitor mass and
radius. These are useful for interpretation of SNe light curves
during the first day, which can teach us about properties of the
progenitor star and potentially to lead to its identification.
Additionally, it can be used to evaluate the effect of the early
emission (e.g., ionization of the circum burst medium), in case that
its detection is missed, on the environment at the SN vicinity.
Finally it is useful for planning targeted searches of shock
breakouts of various SNe types. The theory we discuss here can be
also applied in some cases to shock breakout from non-SN stellar
explosions. For example, the explosion of solar-like star by tidal
forces in the vicinity of a super-massive black hole discussed
recently by \cite{Guillochon09}.

The main conclusions based on our analysis are:
\begin{itemize}
  \item It was shown that shock breakout radiation from WDs, WRs
  and some BSGs is out of thermal equilibrium \citep{Katz09}.
  We show that it typically remains out of thermal equilibrium
  throughout the planar phase and until the early
spherical phase. In SN from these compact progenitors the observed
temperature at this time is significantly higher than the one
obtained when thermal equilibrium is assumed. The observed
temperature falls as $t^{-\alpha}$, where $1/3<\alpha<2/3$, during
the planar phase, and once the evolution becomes spherical it
plunges down (roughly as $t^{-2}$) until we observe radiation that
is in thermal equilibrium at the source.
\item Breakouts from RSGs and some BSGs are in thermal equilibrium.
The flux at frequencies below $T_{obs}$ (e.g., optical/UV) starts
with a bright initial pulse and then it decays during the planar
phase reaching a minimum at $t_s$. The flux is rising during the
planar phase reaching a second maximum when $T_{obs}$ falls into the
observed frequency.
  \item In cases where the radiation is in thermal equilibrium at
the source, the location of the thermalization depth, $r_{cl}$ is
not trivial \citep[e.g.,][]{Ensman92}. The assumptions used in some
previous analytic calculations, such as $r_{cl}=r(\tau=1)$
\citep[e.g., ][]{Chevalier08} or $r_{cl}=\rh$ \citep[e.g.,
][]{Piro09} are incorrect. Instead, in this case
$\rh<r_{cl}<r(\tau=1)$ and it satisfies $r_{cl} =r({\eta=1})$, where
$T_{obs} = (L\tau_{cl}/4\pi \sigma r_{cl}^{~2})^{1/4}$.
\item The initial pulse, which is smeared by light travel-time
and asphericity, contains emission from gas at different
temperatures. The resulting combined spectrum is not a black-body.
Instead, below the peak of $\nu F_\nu$ the spectrum is a power-law
$F_\nu \propto \nu^{-\beta}$ where $0<\beta<1/2$. If asphericity is
negligible then this effect is negligible in RSGs, minor in BSGs and
significant in WRs, where a power-law $F_\nu \propto \nu^{-\beta}$,
ranges over one or two orders of magnitude in $\nu$.
  \item  The bolometric luminosity is dictated by the balance between
  the radiation diffusion time scale and the dynamical time scale
  \citep[e.g.,][]{Chevalier92}. Namely, the source of the observed energy
  at any time is the shell where these two timescales are
  comparable. Since the photons dominate the heat capacity, the bolometric luminosity does not
  depend on the photon-gas coupling,  and it evolves
  independently of the thermal coupling.
  \item The bolometric light curve and spectral evolution of the initial pulse
depends on the explosion asphericity and wind transparency. A
``bare" spherical explosion has a unique and therefore identifiable
light curve. It shows multiple time scales as well as spectral
evolution during the initial pulse. It rises quickly, over $t_0$,
 and lasts for $R_*/c$. A mildly opaque wind
($1<\tau_w<c/\vt$) results in a pulse of characteristic time $\tau_w
R_*/c$ during which the spectrum does not evolve significantly. A
``bare" aspherical explosion produces a time evolving spectrum,
whose light curve and spectral evolution depends on the asphericity
details. Much can be learned from observation of the initial pulse!
  \item The initial pulse of a RSG may release $\sim 3 \cdot
10^{46}$ erg over a duration of $\sim 1000$ sec in soft X-rays
$\approx 0.2$ keV. A BSG shock breakout releases a comparable amount
of energy in harder X-rays over $\sim 100$ sec. The breakout from a
WR releases $\sim 10^{45}$ erg within $\sim 10$ sec in the form of
hard x-rays or soft $\gamma$-rays. Thus, not only WR stars, but also
breakouts from BSGs and potentially RSGs are predicted to produce
strong X-ray flares that can be detected by X-ray telescopes to
large distances. WR breakouts may also be hard enough to be
detectable by current soft gamma-ray detectors to a distance of
$\sim 10$ Mpc, and mimic a single pulse long gamma-ray burst.
  \item The early emission from a type Ia SNe is in $\gamma$-rays.
It is detectable within our galaxy by current detectors and may
mimic a very short ($10-20$ ms) gamma-ray burst.
\end{itemize}

We thank Carlos Badenes, Orly Gnat, Chris Hirata, Amir Levinson,
Tsvi Piran, Sterl Phinney  and Amiel Sternberg for helpful
discussions. E.N. was partially supported by the Israel Science
Foundation (grant No. 174/08) and by an IRG grant. R.S. was
partially supported by ERC and IRG grants, and a Packard
Fellowships.
\renewcommand{\theequation}{A -\arabic{equation}}
\setcounter{equation}{0}  
\section*{APPENDIX A- parameters of the breakout shell}  

Here we give a short list of values of various physical parameters
of the \bo shell at the time of breakout. We consider three
different progenitors of core collapse SNe - RSG (n=1.5, $\kappa =
0.34 {\rm~cm^2 g^{-1}}$), BSG (n=3, $\kappa = 0.34 {\rm~cm^2
g^{-1}}$) and WR (n=3, $\kappa = 0.2 {\rm~cm^2 g^{-1}}$).

We approximate the density profile near the star surface as:
\begin{equation}\label{EQ rho}
    \rho \approx \rho_* \left(\frac{d_i}{R_*}\right)^n
\end{equation}
where $\rho_*=M_*/R_*^3$. There is a correction factor of order
unity to eq. \ref{EQ rho}, which depends on the stellar structure,
but the results are practically insensitive to this factor
\citep{Calzavara04}. The properties at different locations near the
stellar surface are \:
\begin{equation}
    m=\frac{4\pi
    R_*^3\rho_*}{n+1}\left(\frac{\rho}{\rho_*}\right)^\frac{n+1}{n}
\end{equation}
\begin{equation}
    \tau=    \frac{\kappa
    R_*\rho_*}{n+1}\left(\frac{\rho}{\rho_*}\right)^\frac{n+1}{n}
\end{equation}
\begin{equation}
    v = 1800 {\rm~ km/s~} \left(\frac{E_{51}}{M_{15}}\right)^{1/2} \left(\frac{\rho}{\rho_*}\right)^{-0.19}
\end{equation}
Note that we ignore the difference between the ejected mass (used
,e.g., by \citealt{MatznerMcKee99})  and the total stellar mass. The
following properties of the \bo shell at the time of breakout are
found by requiring $\tau=c/\vt$.
\begin{equation}
    \mt \approx \left\{\begin{array}{c}
                        10^{-3} ~M_\odot~ M_{15}^{0.43} R_{500}^{2.26} E_{51}^{-0.56}~~~~~{\rm(RSG)}\\
                        3 \cdot 10^{-6} ~M_\odot~ M_{15}^{0.41} R_{50}^{2.34}E_{51}^{-0.58}~~~~~{\rm(BSG)}\\
                        3 \cdot 10^{-8}~M_\odot~ M_{15}^{0.41} R_{5}^{2.34} E_{51}^{-0.58}~~~~~{\rm(WR)}
                       \end{array}\right.
\end{equation}
\begin{equation}
\rho_0 \approx \left\{\begin{array}{c}
                        6 \cdot 10^{-10} {\rm ~g/cm^3~} M_{15}^{0.67} R_{500}^{-1.64} E_{51}^{-0.3}~~~~~{\rm(RSG)}\\
                        2 \cdot 10^{-9} {\rm ~g/cm^3~} M_{15}^{0.56} R_{50}^{-1.25}E_{51}^{-0.44}~~~~~{\rm(BSG)}\\
                        6 \cdot 10^{-8} {\rm ~g/cm^3~} M_{15}^{0.56} R_{5}^{-1.25} E_{51}^{-0.44}~~~~~{\rm(WR)}
                       \end{array}\right.
\end{equation}
\begin{equation}
    \vt  \approx \left\{\begin{array}{c}
                        7000 {\rm ~km/s~} M_{15}^{-0.43} R_{500}^{-0.26} E_{51}^{0.56}~~~~~{\rm(RSG)}\\
                        20,000 {\rm ~km/s~} M_{15}^{-0.41} R_{50}^{-0.33}E_{51}^{0.58}~~~~~{\rm(BSG)}\\
                        40,000 {\rm ~km/s~} M_{15}^{-0.41} R_{5}^{-0.33} E_{51}^{0.58}~~~~~{\rm(WR)}
                        \end{array}\right.
\end{equation}
\begin{equation}
    E_0 \approx \left\{\begin{array}{c}
                        9 \cdot 10^{47} {\rm ~erg~} M_{15}^{-0.43} R_{500}^{1.74} E_{51}^{0.56}~~~~~{\rm(RSG)}\\
                        3 \cdot 10^{46} {\rm ~erg~} M_{15}^{-0.41} R_{50}^{1.66}E_{51}^{0.58}~~~~~{\rm(BSG)}\\
                        9 \cdot 10^{44} {\rm ~erg~} M_{15}^{-0.41} R_{5}^{1.66} E_{51}^{0.58}~~~~~{\rm(WR)}
                       \end{array}\right.
\end{equation}
Note that $E_0$ is the internal energy in the \bo shell at $t_0$,
while $E_{51}$ is the total explosion energy in units of $10^{51}$
erg.
\begin{equation}
    \tau_0 \approx \left\{\begin{array}{c}
                        50  ~M_{15}^{0.43} R_{500}^{0.26} E_{51}^{-0.56}~~~~~{\rm(RSG)}\\
                        15  ~M_{15}^{0.41} R_{50}^{0.33}E_{51}^{-0.58}~~~~~{\rm(BSG)}\\
                        8  ~M_{15}^{0.41} R_{5}^{0.33} E_{51}^{-0.58}~~~~~{\rm(WR)}
                       \end{array}\right.
\end{equation}
\begin{equation}
    \frac{d_0}{R_*} \approx \left\{\begin{array}{c}
                        5.8 \cdot 10^{-3} ~M_{15}^{-0.22} R_{500}^{0.9} E_{51}^{-0.22}~~~~~{\rm(RSG)}\\
                        6.1 \cdot 10^{-3}   ~M_{15}^{-0.15} R_{50}^{0.58} E_{51}^{-0.15}~~~~~{\rm(BSG)}\\
                        1.9 \cdot 10^{-3}  ~M_{15}^{-0.15} R_{5}^{0.58} E_{51}^{-0.15}~~~~~{\rm(WR)}
                       \end{array}\right.
\end{equation}
\begin{equation}
    \tdi \approx \left\{\begin{array}{c}
                        300 {\rm ~s~} M_{15}^{0.21} R_{500}^{2.16} E_{51}^{-0.79}~~~~~{\rm(RSG)}\\
                        10  {\rm ~s~} M_{15}^{0.27} R_{50}^{1.91} E_{51}^{-0.73}~~~~~{\rm(BSG)}\\
                        0.2  {\rm ~s~} M_{15}^{0.27} R_{5}^{1.91} E_{51}^{-0.73}~~~~~{\rm(WR)}
                       \end{array}\right.
\end{equation}
\begin{equation}
    \frac{c \tdi}{R_*} \approx \left\{\begin{array}{c}
                        0.25 ~M_{15}^{0.21} R_{500}^{1.16} E_{51}^{-0.79}~~~~~{\rm(RSG)}\\
                        0.09   ~M_{15}^{0.27} R_{50}^{0.91}E_{51}^{-0.73}~~~~~{\rm(BSG)}\\
                        0.014   ~M_{15}^{0.27} R_{5}^{0.91}E_{51}^{-0.73}~~~~~{\rm(WR)}
                       \end{array}\right.
\end{equation}
\begin{equation}
    t_s \approx \left\{\begin{array}{c}
                        14 {\rm ~hr~} M_{15}^{0.43} R_{500}^{1.26} E_{51}^{-0.56}~~~~~{\rm(RSG)}\\
                        0.5 {\rm ~hr~}   M_{15}^{0.41} R_{50}^{1.33} E_{51}^{-0.58}~~~~~{\rm(BSG)}\\
                        90 {\rm ~s~}   M_{15}^{0.41} R_{5}^{1.33} E_{51}^{-0.58}~~~~~{\rm(WR)}
                       \end{array}\right.
\end{equation}
\begin{equation}
   \eta_0 \approx \left\{\begin{array}{c}
                        0.06~ M_{15}^{-1.72} R_{500}^{-0.76} E_{51}^{2.16}~~~~~{\rm(RSG)}\\
                        2~ M_{15}^{-1.63} R_{50}^{-1.1} E_{51}^{2.24}~~~~~{\rm(BSG)}\\
                        24~ M_{15}^{-1.63} R_{5}^{-1.1} E_{51}^{2.24}~~~~~{\rm(WR)}
                       \end{array}\right.
\end{equation}
\begin{equation}
   T_{obs,0} \approx \left\{\begin{array}{lc}
                        25 {\rm ~eV~} M_{15}^{-0.3}R_{500}^{-0.65} E_{51}^{0.5}&~~~~~{\rm(RSG)}\\
                        80  {\rm ~eV~} M_{25}^{-0.28} R_{70}^{-0.62} E_{51}^{0.48}& ~~~~~{\rm(BSG~ thermal)}\\
                        700 {\rm ~eV~} M_{10}^{-1.2} R_{20}^{-1.1} E_{51}^{1.7}& ~~~~~{\rm(BSG ~nonthermal)}\\
                        2 {\rm ~keV~} M_{15}^{-1.7} R_{5}^{-1.5} E_{51}^{1.8}&~~~~~{\rm(WR)}
                       \end{array}\right.
\end{equation}
Note that $T_{obs,0}$ is the observed temperature at the time of
breakout and not necessarily the temperature of the \bo shell at
that time. The dependence of $T_{obs,0}$ in the nonthermal cases
(BSG ~nonthermal and WR) on $M_*$, $R_*$ and $E_{51}$ is
approximated and calculated numerically. The reason is that the
value of the logarithmic Comptonization factor $\xi$ has a
non-power-law dependence on the explosion parameters. The power law
indices we provide are accurate to within $\pm 0.5$ when $R_*$ and
$M_*$ are larger or smaller by up to a factor of about 3, than our
canonical values and $E_{51}=1-2$. In order to obtain a more
accurate evaluation of nonthermal $T_{obs,0}$ when $M_*$, $R_*$ and
$E_{51}$ differ from our canonical values,  $7\rho_0$ (the factor of
$7$ is due to the shock compression) should be plugged into equation
\ref{EQ ymax} ($y_{max})$ and then equation \ref{EQ Teta1} should be
solved numerically (using $\eta_0$).

\renewcommand{\theequation}{B-\arabic{equation}}
\setcounter{equation}{0}  
\section*{APPENDIX B - Electron-photon temperature coupling}
Our temperature calculations (in particular equation \ref{EQ Teta1})
assume that in the relevant shells the electron temperature follows
the photon temperature at all time. Here we examine the validity of
this assumption.

Consider first a shell that is in thermal equilibrium at the time of
breakout (i.e., $\eta_i <1$). As discussed in section \ref{SEC
thermal coupling} when $\eta<1$ a typical photon is absorbed at
least once within the available time ($\min\{t,t_d\}$). Therefore,
since photons dominate the energy density, the electron temperature
is coupled to the radiation while $\eta<1$. Now, if in a shell
$\eta_i <1$ then by the time that  $\eta>1$ that shell does not
affect the observed temperature anymore since at this point it is
external to the \lum shell. Therefore in such shells our temperature
coupling coupling holds while relevant.

Next consider a shell that is initially out of thermal equilibrium
(i.e., $\eta_i> 1$). Here, absorbtion do not play a significant role
and electrons follow the photons temperature if the rate at which
they lose energy (electron cooling rate), mostly via free-free
emission, is slower than the rate at which they gain energy via
Compton scattering with typical photons (electron heating rate). The
coupling in shells that are out of thermal equilibrium is important
mostly during the planar phase, since during the spherical phase the
color shell is quickly receding inward to shells that are in thermal
equilibrium. Since during the planar phase only the \bo shell is
observed (it is also the \lum shell and the color shell when
$\etah>1$) we consider only the coupling in this shell. Below we
calculate the heating and cooling rates in the \bo shell that is out
of thermal equilibrium during the planar phase.

The electron heating term is:
\begin{equation}
    \dot{e}_{heat}=\frac{c}{l}\frac{n_{ph}}{n_e}\frac{3kT}{m_ec^2}3kT
\end{equation}
where here we denote number densities as $n$ with different
subscripts, not to be confused with the stellar envelope structure
power-law index. $e$ is the energy of a single electron,
$l=1/n_e\sigma_T$ is the photons mean free pass, $n_{ph}$ is the
photon density, $n_e$ is the electron density and we take $3kT$
photons as responsible for most of the heating. Just behind the
shock $n_{ph,i} \approx \frac{n_u m_p v_0^2}{kT_i}$ where $n_u$ is
the upstream density (compressed by a factor of 7 in the shock,
i.e., $n_{e,i}=7n_u$), implying:
\begin{equation}
    \dot{e}_{heat}\approx c \sigma_Tn_e\frac{m_p v^2}{m_ec^2}
    \frac{E}{E_i}kT
\end{equation}
where $E$ is the total energy in the radiation and $E_i$ is its
initial value. During the planar phase $E/E_i \propto t^{-1/3}$.

The main cooling process is free-free emission:
\begin{equation}
    \dot{e}_{cool} \approx \alpha \sigma_T c n_e  \sqrt{m_ec^2kT}
\end{equation}
where $\alpha$ is the fine structure constant and we neglect the
logarithmic factor $\xi$.  We define the heating to cooling ratio:
\begin{equation}
    \chi\equiv\frac{\dot{e}_{heat}}{\dot{e}_{cool}}\approx  4 \left(\frac{v_0}{10^4~km/s}
    \right)^{2}\left(\frac{kT}{100~eV}\right)^{1/2}
    \left(\frac{t}{t_0}\right)^{-1/3}
\end{equation}

If $\chi>1$ then electrons track the photons temperature, while if
$\chi<1$ then Compton scattering is not enough to keep the electrons
at the photons temperature. Thus, If $\chi<1$ and in addition $\eta
> 1$ in the \bo shell during the planar phase, then not Compton
scattering nor free-free absorbtion can heat the electrons fast
enough and our assumption of a single temperature fails. Using
equations \ref{EQ eta}, \ref{EQ eta planar} and $kT_{BB,0} \approx
(v_0\rho_0/a)^{1/4}$ we find that while $\eta>1$ then in the \bo
shell:
\begin{equation}
    \chi \approx 5 \eta^{8/5} \left(\frac{\rho_0}{10^{-10}~g/cm^{-3}}\right)^{1/5} \left(\frac{t}{t_0}\right)^{-2/5}.
\end{equation}
Thus, our assumption of a single electron-photon temperature is
valid during the breakout. Moreover, $\chi$ depends strongly on
$\eta$ and only weakly on $t$ (e.g., in a typical BSG,
$(t_s/t_0)^{2/5} \approx 8$). Therefore, the request $\chi>1$ while
$\eta > 1$ may be at most marginally violated in the \bo shell
during the planar phase, and only in a narrow range of the parameter
phase space. Note that even if the electron temperature in the
breakout shell decouples from that of the photons during the planar
phase the modifications to the observed light curve are minor. The
bolometric luminosity is of course not affected while the
temperature in the \bo shell is not driven anymore towards thermal
equilibrium and it starts dropping only via adiabatic cooling, e.g.,
$T_{obs} \propto t^{-1/3}$ (instead of a slightly faster decay).
Later, the temperature evolution during the early spherical phase is
also slightly modified. Nevertheless, the general behavior, that the
temperature drops very sharply at $t_s<t$ until it approaches
$\Th_{BB}$ and merges with the thermal equilibrium evolution at $t
\approx t_2$, is unchanged.

Therefore, we conclude that our assumption that the electrons and
photons temperatures are similar generally holds in the shells that
determine the temperature evolution, and that even if it is violated
over a small region of the relevant parameter phase space the
modifications are small.

\section*{APPENDIX C - Glossary of main symbols and notations}

\begin{itemize}
\renewcommand{\labelitemi}{$\cdot$}
\item $t$: time since breakout
\item $r$: radius
\item $v$: velocity
\item $m(r)$: mass at radius larger than $r$
\item $\rho$: mass density
\item $d$: shell width
\item $\tau$: optical depth
\item $E$: internal energy (not to be confused with $E_{51}$)
\item $t_d$: diffusion time
\item $\epsilon$: energy density
\item $\eta$: Thermal coupling coefficient. Defined in equation \ref{EQ eta
Def}
\item $\xi$: logarithmic Comptonization term. Defined in equation
\ref{EQ xi}
\item $L$: observed luminosity
\item $T_{obs}$: observed temperature (often denoted as color temperature).
Defined as the typical photon energy.
\item \bo shell: The shell where the shock breaks out (the shock width
is comparable to the shell width).
\item \lum shell: The shell that generates the observed luminosity.
In this shell the diffusion time equals to the dynamical time, t.
\item color shell: The shell where the observed temperature is
determined. This shell coincides with the thermalization depth when
the observed radiation is in thermal equilibrium and with the \lum
shell when it is not.

\item For any quantity $x$, we use the following subscripts and
superscripts\\
$x_i$: Initial value (after shock crossing) of a
shell.\\
$x_0$:  value at the \bo shell at the time of breakout.\\
$\widehat{x}$: value at the \lum shell.\\
$x_{cl}$: value at the color shell.
\item $t_0$: Duration of shock breakout. Also, dynamical time, diffusion time,
and shock crossing time of the \bo shell at the time of breakout.
\item $t_1$: The first time where the observed temperature is in thermal equilibrium if it was
not in thermal equilibrium from the beginning. Thermal equilibrium
was achieved by photons produced at earlier time. Defined in
equation \ref{EQ t1}
\item $t_2$: The first time where the observed temperature is in thermal equilibrium,
which is achieved by photons produced at $t_2$, if it was not in
thermal equilibrium from the beginning. Defined in equation \ref{EQ
t2}
\item $t_s$: The transition time from planar to spherical geometry
($=R_*/v_0$).
\item $T_{obs,0}$: observed temperature at $t_0$. Equal to the temperature of the \bo shell
if it is out of thermal equilibrium. Otherwise it is lower than the
\bo shell temperature.
\item $T_{BB}$: Thermal equilibrium temperature appropriate for a given energy density by Boltzmann's law.
Defined in equation \ref{EQ TBB_def}.
\item $R_*$: stelar radius
\item $M_*$: stelar mass
\item $E_{51}$: explosion energy in units of $10^{51}$ erg
\item $n$: power-law index describing the pre-explosion stellar density
profile near the edge.
\end{itemize}


\end{document}